\documentclass[aps,pre,twocolumn,showpacs,superscriptaddress,groupedaddress]{revtex4-1}
\pdfoutput=1
\usepackage{hyperref}
\usepackage{epsfig}
\usepackage{amsmath}
\usepackage{graphicx}
\usepackage{wrapfig}
\usepackage{dcolumn}
\usepackage{bm}
\usepackage{amssymb}
\usepackage{titlesec}
\usepackage{mathtools}
\usepackage{relsize}
\usepackage{cleveref}
\usepackage[usenames,dvipsnames]{color}

\begin{document}
\title{Stability of milling patterns in self-propelled swarms on surfaces}
\author{Jason Hindes$^{1}$, Victoria Edwards$^{1}$, Sayomi Kamimoto$^{2}$, George Stantchev$^{1}$, and Ira B. Schwartz$^{1}$}
\affiliation{$^{1}$U.S. Naval Research Laboratory, Washington, DC 20375, USA}
\affiliation{$^{2}$Department of Mathematics, George Mason University, Fairfax Virginia, 22030, USA}

\begin{abstract}
  In some physical and biological swarms, agents effectively move and interact along curved surfaces.
  The associated constraints and symmetries can affect collective-motion patterns, but little is known about pattern stability in the presence of surface curvature. 
  To make progress, we construct a general model for self-propelled swarms moving on surfaces
  using Lagrangian mechanics. We find that the combination of self-propulsion, friction, mutual attraction, and surface curvature produce milling patterns where each agent in a swarm
  oscillates on a limit cycle, with different agents splayed along the cycle such that the swarm's center-of-mass remains stationary.
  In general, such patterns loose stability when mutual attraction is insufficient to overcome the constraint of curvature, and we uncover two broad classes 
  of stationary milling-state bifurcations. In the first, a spatially periodic mode undergoes a Hopf bifurcation as curvature is increased which results in unstable spatiotemporal oscillations. 
  This generic bifurcation is analyzed for the sphere and demonstrated numerically for several surfaces. In the second, a saddle-node-of-periodic-orbits 
  occurs in which stable and unstable milling states collide and annihilate. The latter is analyzed for milling states on cylindrical surfaces. 
  Our results contribute to the general understanding of swarm pattern-formation and stability in the presence of surface curvature, and may aid in designing robotic swarms that can be controlled to move over complex surfaces. 
\end{abstract}
\maketitle

\section{\label{sec:Intro} INTRODUCTION}
Swarming occurs when emergent spatiotemporal patterns are produced from the interaction of large collectives of coupled mobile agents with limited dynamics and simple rules.
Examples have been discovered over a wide range of space and time scales in nature including: colonies of bacteria and bees\cite{Polezhaev,Li_Sayed_2012}, swarms of ants and locusts\cite{Theraulaz2002,Topaz2012}, schools of fish\cite{Couzin2013, Calovi2014, Cavagna2015}, flocks of starlings and jackdaws\cite{Leonard2013, Ballerini08,Ouellette2019}, and crowds of people\cite{Rio_Warren_2014}.
As a consequence, much work in biophysics has focused on studying simple models to understand general swarm pattern-formation\cite{Vicsek,Marchetti,Aldana,PhysRevX.9.011002}, and the nonequilibirum statistics of active-matter systems\cite{Solon2015,PhysRevLett.117.038103,PhysRevLett.121.178001,PhysRevLett.122.258001}. 

In general, natural swarms are robust to the
loss of individual agents and communication links, and are dynamically responsive to complex, changing environments.
Because of these useful properties, there is great interest in engineering multi-robot systems that can imitate nature's ability to produce robust collective behavior from simple components. Research in robotic swarming has focused on developing control laws that govern individual robot motion within a group so that stable group-dynamics emerges\cite{Desai01, Jadbabaie03, Tanner03b, Tanner03a,  Gazi05, Tanner07}. In addition to pattern formation, robotic swarms have been proposed for solving a variety of cooperative-dynamics tasks including mapping\cite{Ramachandran2018}, resource allocation \cite{Li17, Berman07, Hsieh2008}, and leader following\cite{Wiech2018}.

Swarms in three-spatial dimensions can be subject to the geometry and curvature of surfaces on which they move.  
Examples range from embryonic development\cite{Science2008} and corneal growth\cite{CornealGrowth} to the collective dynamics of active particles
confined to droplets\cite{Topology2014,NematicShells} and robotic consensus
and control on manifolds\cite{ConsensusOnSphere} (e.g., tracking and sensing
on a surface with known geometry). Yet, most of the known theoretical results
concerning swarms on surfaces pertain to novel steady-state patterns due to curvature, equilibrium statistics,
and relaxation properties\cite{PhysRevE.91.022306,Li2015,PhysRevE.97.052615,Janssen2017,Riemannian2018,PhysRevE.97.052605}. 
Despite extensive work on general swarm stability\cite{GaziStabilityBook,AlbiStability2014,PhysRevE.101.042202}, to our knowledge little is known about the detailed bifurcation structure of collective swarming patterns on various curved surfaces.

Here, we analyze stationary, milling patterns of self-propelled swarms
on surfaces with simple attractive interactions. We find two types of
bifurcations that appear as surface curvature is increased: a generalized Hopf
bifurcation where spatially-periodic modes introduce unstable oscillations at
a definite frequency, and a saddle-node-of-periodic-orbits (SNpo) bifurcation
where stable and unstable milling states coalesce. Our results shed light on
how surface curvature destabilizes swarming patterns in both generic and
geometry-specific ways, and provides insight into how patterns emerge on
different surfaces. In addition to general understanding, the techniques used to reveal bifurcation structure may aid in designing robotic swarms to move cooperatively over complex surfaces and terrains.   

To begin, let us consider a swarm of self-propelled agents interacting through conservative forces -- the simplest model for position-dependent interactions. In addition to such forces, each agent is acted upon by activation-dissipation forces that 
depend on its velocity, $\bold{A}_{l}\!=\!\mu(v_{l})\bold{v}_{l}/v_{l}$, where $\bold{v}_{l}$ is 
the velocity of the $l$th agent, $v_{l}$ is its speed, and $\mu(v_{l})$ is a nonlinear function to be specified\cite{Levine,Erdmann,Minguzzi}. The latter is a consistent feature of active matter systems, since the ``friction coefficient", $\mu(v_{l})$, can be both positive and negative, due to input energy from an agent's environment\cite{Levine,Erdmann,DOrsagna,Minguzzi} (or onboard motor in the case of mobile robots\cite{Edwards2019,Szwaykowska2016}). 

We are interested in studying a swarm's dynamics in generalized
curvilinear coordinates, $q^{k}_{l}$, in which Lagrange's equations-of-motion take the covariant form
\begin{align}
\label{eq:GeneralLagrange}
\frac{d}{dt}\frac{\partial L}{\partial \dot{q}^{k}_{l}} - \frac{\partial L}{\partial q^{k}_{l}} + \frac{\partial F}{\partial \dot{q}^{k}_{l}} =0, 
\end{align} 
where $L$ is the swarm's Lagrangian, $F\!=\!\sum_{l}\int_{0}^{v_{l}}\!\mu_{l}(v_{l})dv_{l}$ is the Rayleigh dissipation (activation) function\cite{Minguzzi}, and dots denote time derivatives. Equation(\ref{eq:GeneralLagrange}) is standard for Lagrangian mechanics with dissipation -- the important distinction being the possibility of effectively negative dissipation for active agents. The Lagrangian approach is expected to significantly simplify the description and analysis of collective-motion states in swarms, especially in the presence of symmetries and generalized constraints, as in classical mechanics problems. Moreover, it does not assume an over-damped limit for the agent dynamics\cite{Riemannian2018,PhysRevE.91.022306}.     

For simplicity, in this paper we take 
\begin{align}
\label{eq:Dissipation}
&F=-\sum_{n}\frac{v_{n}^{2}}{2}\!\Big(\alpha -\frac{\beta}{2}v_{n}^{2}\Big), \\
\label{eq:Lagrangian}
&L=m\sum_{n}\frac{v_{n}^{2}}{2} -\frac{a}{4N}\sum_{n,j}|\bold{r}_{n}-\bold{r}_{j}|^{2},
\end{align}
as in many works on swarms of self-propelled agents in Cartesian coordinates\cite{Levine,Erdmann,DOrsagna,F1,Romero2012} (which we recover as a special case).  In Eqs. (\ref{eq:Dissipation}-\ref{eq:Lagrangian}) $m$ is the mass of each agent, $\alpha$ is a self-propulsion constant, $\beta$ is a friction constant, $a$ is a coupling constant, $N$ is the number of agents, and $\bold{r}_{j}\!=\!\left(x_{j},y_{j},z_{j}\right)$ is the position-vector in Cartesian coordinates for the $j$th agent in three spatial dimensions. Note that Eq.(\ref{eq:Lagrangian}) implies that the pairwise interaction force between two agents, $n$ and $j$ is linear in $\bold{r}_{n}\!-\!\bold{r}_{j}$ and global, though these assumptions can be relaxed\cite{Szwaykowska2016,J1}. In addition, note that the spring-like potential in Eq.(\ref{eq:Lagrangian}) is defined in terms of the pairwise distance between agents in Euclidean space, rather than along, e.g., geodesics \cite{Osborne2013}. This is done for the sake of simplicity and to allow for a proof of principle in the weak mean-curvature limit. Beyond theory, we note that Cartesian versions of Eqs.(\ref{eq:Dissipation}-\ref{eq:Lagrangian}) have been implemented in experiments with several robotics platforms including autonomous cars, boats, and quad-rotors\cite{Szwaykowska2016,Edwards2019}. 

In what follows we study the stability of swarms on surfaces using the above formalism. 
In Sec.\ref{sec:Cycle}, we define the milling patterns of interest as swarm limit-cycles, and compute their properties on several surfaces.
In Sec.\ref{sec:Stability}, we analyze the stability of limit cycles by tracking which Floquet multipliers cross the unit circle under parameter variation. 
Two destabilization patterns are observed as surface curvature is increased. In the first, two milling-state multipliers cross in a Hopf bifurcation of spatially-periodic modes. 
In the second, many multipliers cross simultaneously, resulting in a SNpo bifurcation. The first is discussed in Sec.\ref{sec:periodic}, while the second is discussed in Sec.\ref{sec:single}. 
Throughout, predictions are compared to simulations. Sec.\ref{sec:Conclusion} provides a summary and further discussion. 
  
\section{\label{sec:Cycle} LIMIT-CYCLE MILLING}
We are interested in the dynamics of Eqs.(\ref{eq:GeneralLagrange}-\ref{eq:Lagrangian}) when agents are constrained to surfaces, e.g., when a particular curvilinear coordinate is constant. Classic example surfaces considered in detail are: the sphere of radius~$r$ $\left(x_{l},y_{l},z_{l}\right)_{s}\!=\!r\left(\sin\theta_{l}\!\cos\phi_{l}, \sin\theta_{l}\!\sin\phi_{l}, \cos\theta_{l}\right)$; the cylinder of radius $\rho$, $\left(x_{l},y_{l},z_{l}\right)_{c}\!=\!\left(\rho\cos\phi_{l},\rho\sin\phi_{l},z_{l}\right)$, and the torus of radii $b$ and $c$, $\left(x_{l},y_{l},z_{l}\right)_{t}\!=\!\left((c+b\cos\theta_{l})\!\cos\phi_{l},(c+b\cos\theta_{l})\!\sin\phi_{l},b\sin\theta_{l}\right)$. However, our approach can be generally applied to any coordinate surface.

Substituting these curvilinear coordinates into Eqs.(\ref{eq:GeneralLagrange}-\ref{eq:Lagrangian}) gives the example equations-of-motion for swarming on a sphere,
\begin{align}
\label{eq:Sph1}
&\ddot{\phi}_{l}\sin\theta_{l} - \dot{\phi}_{l}\sin\theta_{l}\Big[\alpha-\!\beta r^{2}\!\big(\dot{\theta}_{l}^{2} + \dot{\phi}_{l}^{2}\sin^{2}\theta_{l}\big)\!\Big]-2\dot{\phi}_{l}\dot{\theta}_{l}\!\cos\theta_{l} \;\nonumber\\
&- \frac{a}{N}\!\sum_{j}\sin\theta_{j}\!\sin(\phi_{j}\!-\phi_{l})  = 0,\\
\label{eq:Sph2}
&\ddot{\theta}_{l} - \dot{\theta}_{l}\Big[\alpha-\!\beta r^{2}\!\big(\dot{\theta_{l}}^{2} + \dot{\phi}_{l}^{2}\sin^{2}\dot{\theta}_{l}\big)\!\Big] - \dot{\phi}_{l}^{2}\sin\theta_{l}\!\cos\theta_{l}\;\; \nonumber \\
&-\frac{a}{N}\!\sum_{j}\!\Big[\!\sin\theta_{j}\!\cos\theta_{l}\!\cos(\phi_{j}\!-\phi_{l})-\cos\theta_{j}\!\sin\theta_{l}\!\Big] = 0,
\end{align} 
and on a cylinder,
\begin{align}
\label{eq:Cyl1}
&\ddot{z}_{l}=\dot{z}_{l}\Big[\alpha-\beta\big(\rho^{2}\dot{\phi_{l}}^{2}+\dot{z}_{l}^{2}\big)\!\Big]+\frac{a}{N}\sum_{j}(z_{j}-z_{l}),\\
\label{eq:Cyl2}
&\ddot{\phi}_{l}=\dot{\phi}_{l}\Big[\alpha-\beta\big(\rho^{2}\dot{\phi_{l}}^{2}+\dot{z}_{l}^{2}\big)\!\Big]+\frac{a}{N}\sum_{j}\sin\!\big(\phi_{j}-\phi_{l}\big),
\end{align}
where we set $m\!=\!1$ (since it entails an overall time-constant only). Similar equations for swarming on a torus can be found in the App.\ref{sec:torus}. We note that the same equations-of-motion can be derived from the Cartesian version of Newton's law for the model considered, by substitution, though this approach is more cumbersome\cite{PhysRevE.101.042202}. 

There are two tendencies in the swarm's dynamics worth noting based on the physics of Eqs.(\ref{eq:GeneralLagrange}-\ref{eq:Lagrangian}). First, when $a\!=\!0$ the activation-dissipation forces drive agents to follow independent geodesics on the surface with equal speed, but no particular direction of motion (see App.\ref{sec:Uncoupled}). Second, agents tend to cluster in space, such that the distance between agents is minimized. The latter is implied by the spring-like interaction forces in Eq.(\ref{eq:Lagrangian}). Together the combined forces produce a variety of collective-motion states depending on parameter values and initial conditions. 

\begin{figure*}[t]
\centering
\includegraphics[scale=0.30]{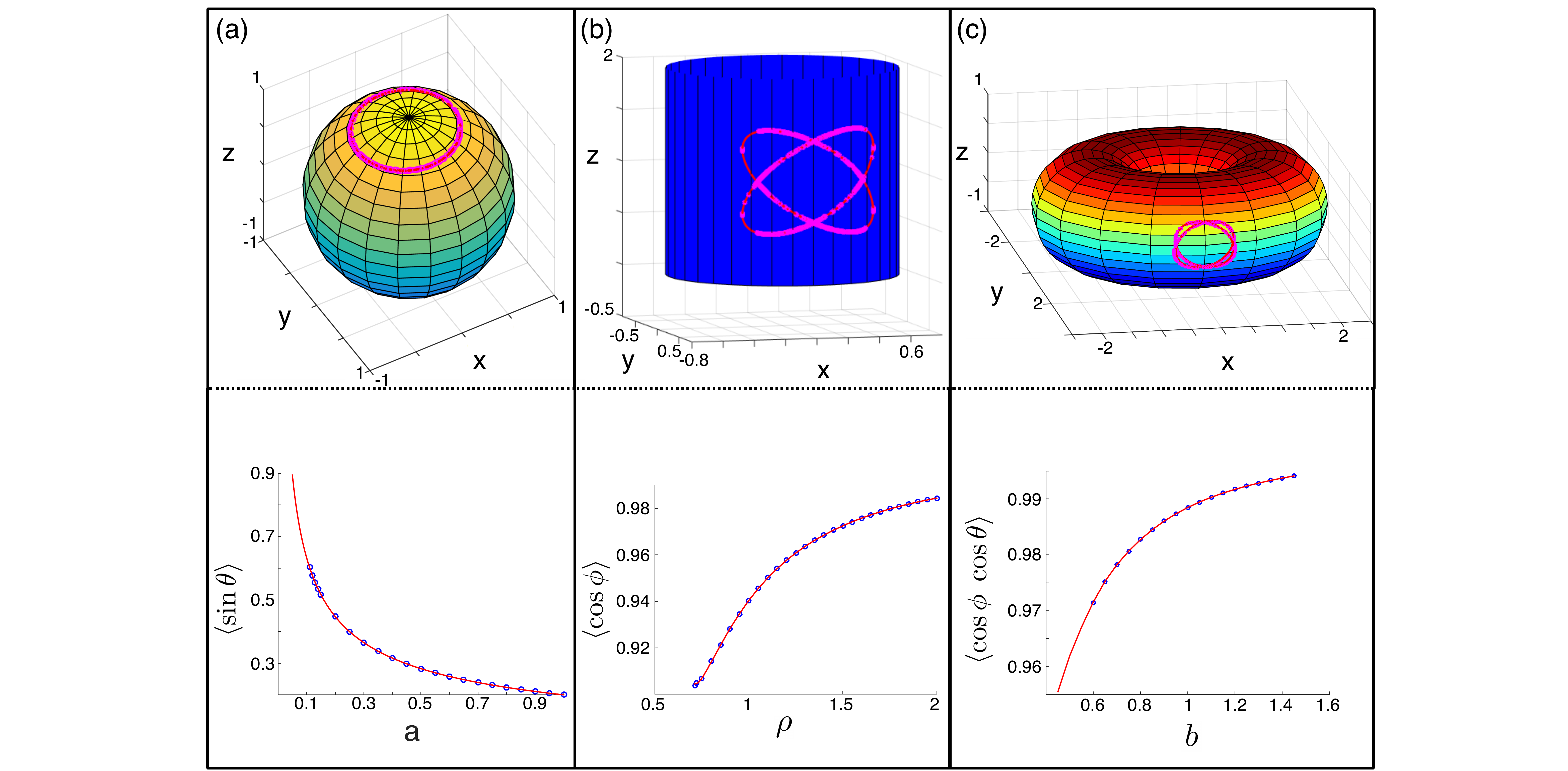}
\caption{(Color online) Milling states on coordinate surfaces. Top panels show time-snapshots from simulations with $N\!=\!600$ agents on the (a) sphere ($r\!=\!1.0,\alpha\!=\!0.2,\beta\!=\!5.0,a\!=\!0.15$), (b) cylinder ($\rho\!=\!0.8,\alpha\!=0.5,\beta\!=\!1,a\!=\!2$), and (c) torus ($b\!=\!0.9,c\!=\!1.5,\alpha\!=\!0.2,\beta\!=\!1.0,a\!=\!1.0$). Agents are drawn with magenta circles and predictions from Eq.(\ref{eq:SelfCons}) with solid red lines. Bottom panels show predictions from Eq.(\ref{eq:SelfCons}) versus a curvature constant for each surface. Remaining parameters are identical to the top panels.}
\label{fig:Patterns}
\end{figure*}

However, in this work we concern ourselves with stationary milling solutions of Eqs.(\ref{eq:GeneralLagrange}-\ref{eq:Lagrangian}) where each agent maintains periodic motion, while the center-of-mass of the swarm, $\bold{R}\!\equiv\!\sum_{j}\bold{r}_{j}/N$, remains constant. A primary reason for studying milling states is that they emerge from broad initial conditions, e.g., spatially uniform distributions of agents over a surface with random initial velocities. In contrast, flocking states require initial alignment of velocities. The periodic motion of milling states occurs on limit cycles (LCs), with agents splayed approximately uniformly at different points along the same LC. The top panels in Fig.\ref{fig:Patterns} show  snap-shots in time of example milling states for large swarms.
 
In general, we can only calculate exact LCs for special cases such as on the sphere. However, we can compute LCs on other surfaces, without resorting to large multi-agent simulations, by using a self-consistency approach in the limit $N\!\gg\!1$. The approach entails trading the sums-over-agents that appear in Eqs.(\ref{eq:Sph1}-\ref{eq:Cyl2}), with time-averages over a single-particle LC. The former requires simulating a 4N-dimensional system, the latter a 4-dimensional system with a properly chosen $\bold{R}$,
\begin{align}
\label{eq:SelfCons}
&\bold{R}=\int_{0}^{T(\bold{R})}\bold{r}^{(LC)}(t;\bold{R})\frac{dt}{T(\bold{R})}. 
\end{align}
The trajectory $\bold{r}^{(LC)}(t;\bold{R})$ is a LC of the single-particle system, and $T(\bold{R})$ is its period. A single-particle system can be found by replacing $\bold{R}$, which appears in the interaction sums in Eqs.(\ref{eq:Sph1}-\ref{eq:Cyl2}), with Eq.(\ref{eq:SelfCons}). For example, substituting Eq.(\ref{eq:SelfCons}) into Eqs.(\ref{eq:Cyl1}-\ref{eq:Cyl2}) we find
\begin{align}
\label{eq:LCcyc1}
&\ddot{z}=\dot{z}\Big[\alpha-\beta\big(\rho^{2}\dot{\phi}^{2}+\dot{z}^{2}\big)\!\Big] -az,\\
\label{eq:LCcyc2}
&\ddot{\phi}=\dot{\phi}\Big[\alpha-\beta\big(\rho^{2}\dot{\phi}^{2}+\dot{z}^{2}\big)\!\Big]-a\sin\phi \left<\cos\phi\right>,
\end{align}
where
\begin{align}
\label{eq:LCcyc3}
&\left<\cos\phi \right> =\int_{0}^{T}\!\cos\phi(t)\;\frac{dt}{T}. 
\end{align}
is the time average of $\cos \phi$ over a LC period.
Note that in Eqs.(\ref{eq:LCcyc1}-\ref{eq:LCcyc3}) we have chosen $\left<z\right>\!=\!\left<\sin\phi\right>\!=\!0$, since these constants simply shift and rotate LCs along the cylinder, and we have dropped the agent subscript, $l$. See App.\ref{sec:torus} for the analogous equations on a torus. 

Limit cycles can be computed numerically by: integrating the single-particle system\cite{Note1}, $\bold{r}(t)$, with an initial choice of $\bold{R}$, determining the period of the LC after transients, and updating the choice of $\bold{R}$ based on a quasi-Newton evaluation of Eq.(\ref{eq:SelfCons}). Such an algorithm can be implemented using numerical-integration software combined with a fixed-point-solver. Example comparisons are shown in Fig.\ref{fig:Patterns} with excellent agreement. In the top panels we plot the LCs in red on each surface compared to large multi-agent simulations shown with magenta circles. The initial conditions for the large multi-agent simulations were uniformly-random spatial distributions over each surface with random velocities. In the bottom panels, we plot solutions of Eq.(\ref{eq:SelfCons}) compared to averages-over-particles in the swarm, shown with blue circles, for a wide range of parameters.

We note that in general when milling states are stable in swarms with attractive interactions, for a given $\bold{R}$ there are two possible stable LCs that are simply reflections of one another, e.g., $\phi\!\rightarrow\!-\phi$ given $\left<z\right>\!=\!\left<y\right>\!=\!0$. Both LCs can be seen clearly in the top panels of Fig.\ref{fig:Patterns}(b-c). As a consequence, a general milling state on a surface is built from some fraction of the agents moving on one LC, while the remaining move on the reflected LC-- the fraction depending on initial conditions. The existence of two LCs also occurs in Cartesian swarms in 2d\cite{DOrsagna,F1,Romero2012}, where reflection simply reveres the angular velocity.    

For the special case of milling on the sphere, LCs are simply circular orbits. Assuming a swarm milling with a uniform distribution of constantly rotating $\phi_{j}$'s all in the same direction, and oriented such that the polar angle is constant, we find
\begin{align}
\label{eq:Ring}
\!\theta_{j}(t)=\sin^{-1}\!\!\Bigg(\!\sqrt{\!\!\frac{\alpha}{\beta a r^{2}}}\Bigg), \;\; \phi_{j}(t)=\frac{2\pi(j-1)}{N}+\sqrt{a}t.                               
\end{align}  
Equation (\ref{eq:Ring}) is easy to check by direct substitution, and compares well to simulations, Fig.\ref{fig:Patterns}(a). Note that for consistency, a LC does not exist on the sphere if $\alpha/[\beta a r^{2}]\!>\!1$. In addition, as with a Cartesian swarm in 2d, the reflected LC on a sphere has the opposite angular velocity, $\dot{\phi}_{j}\!=\!-\sqrt{a}$.
 
\section{\label{sec:Stability} STABILITY}
As an estimate, we expect milling states to change stability when the arc length of the LC is roughly equal to the inverse of the mean surface curvature. If these two quantities are very different, then a periodic-solution should be hard to realize on the surface. The natural period of oscillations for milling scales as $\sim 1/\sqrt{a}$, e.g. Eq.(\ref{eq:Ring}), while the average speed of an agent is $\sqrt{\alpha/\beta}$; the latter is the speed at which self-propulsion and friction forces cancel. See App.\ref{sec:Uncoupled} for more details. Altogether we expect instability to arise approximately when $H\sqrt{\alpha/(\beta a)} \sim 1$, where $H$ is the mean surface curvature. 
  
Beyond this crude estimate, we would like to understand and analyze the stability of milling states quantitatively, and determine if there are any differences in the bifurcations between surfaces. In our stability analysis below, we focus on milling states that correspond to a single LC, where all agents rotate in the same direction, for three reasons: this case persists when repulsive forces are added, the stability of any given milling state has only a weak dependence on the number of agents on each LC, and it is analytically tractable. We begin by analyzing the linear stability of milling states on the sphere, where a complete bifurcation picture can be derived, and compare to numerical Floquet analysis for other surfaces.  
 
To determine the local stability on the sphere we need to understand how small perturbations to Eq.(\ref{eq:Ring}) change in time. Our first step is to substitute a general perturbation, $\theta_{j}(t)\!=\!\sin^{-1}\!\Big[\!\sqrt{\alpha/(\beta a r^{2})}\Big]+\!B_{j}(t)$ and $\phi_{j}(t)\!=\!2\pi(j-1)/N\!+\!\sqrt{a}t +\!A_{j}(t)$, into Eqs.(\ref{eq:Sph1}-\ref{eq:Sph2}) and collect terms to first order in $A_{j}(t)$ and $B_{j}(t)$ (assuming $|A_{j}|\;, |B_{j}| \ll1\;\forall j$). The result is the following linear system of differential equations with constant coefficients -- the latter property is a consequence of our transformation into the proper coordinate system and is what allows for an analytical treatment for milling:
\begin{widetext} 
\begin{align}
\label{eq:LinRing1}
&\ddot{B}_{l}+\!\frac{\alpha}{\beta r^{2}}B_{l}-\!\Bigg(\frac{2\alpha}{\beta \sqrt{a} r^{2}}\sqrt{\tfrac{\beta ar^{2}}{\alpha}\!-\!1}\Bigg)\!\dot{A}_{l}=\frac{\alpha}{\beta r^{2}N}\!\sum_{j}\!\Bigg[\!\Bigg(\!1+\!\Big(\frac{\beta a r^{2}}{\alpha}\!-1\!\Big)\!\cos\!\Big(\!\tfrac{2\pi (j-l)}{N}\!\Big)\!\!\Bigg)B_{j}-\sin\!\Big(\!\tfrac{2\pi (j-l)}{N}\!\Big)\sqrt{\tfrac{\beta ar^{2}}{\alpha}\!-\!1}\cdot A_{j}\Bigg],\\
\label{eq:LinRing2}
&\ddot{A}_{l}+2\alpha\dot{A}_{l}+2\sqrt{a}\sqrt{\tfrac{\beta a r^{2}}{\alpha}\!-\!1}\cdot\!\big[\dot{B}_{l}+\alpha B_{l}\big]=\frac{a}{N}\!\sum_{j}\!\Bigg[\!\sin\!\Big(\!\tfrac{2\pi (j-l)}{N}\!\Big)\sqrt{\tfrac{\beta a r^{2}}{\alpha}\!-\!1}\cdot\!B_{j}+\cos\!\Big(\!\tfrac{2\pi (j-l)}{N}\!\Big)A_{j}\Bigg].                           
\end{align}
\end{widetext}
\subsection{\label{sec:periodic} Spatially periodic modes}
Given the periodicity implied by the splayed phase variables in Eq.(\ref{eq:Ring}), it is natural to look for eigen-solutions of Eqs.(\ref{eq:LinRing1}-\ref{eq:LinRing2}) in terms of the discrete Fourier transforms of $A_{j}(t)$ and $B_{j}(t)$. In fact, by inspection we can see that only the first harmonic survives the summations on the right-hand sides of Eqs.(\ref{eq:LinRing1}-\ref{eq:LinRing2}), because of the sine and cosine terms, and hence we look for the particular solutions: $A_{j}(t)=A\exp(\lambda t +2\pi i(j-1)/N)$ and $B_{j}(t)=B\exp(\lambda t +2\pi i(j-1)/N)$. Substitution gives the following equation for the stability exponent, $\lambda$, of the spatially-periodic modes on the sphere:
\begin{align}
\label{eq:PeriodicMode}
&\Bigg[\lambda^{2}+2\alpha\lambda-\frac{a}{2}\Bigg]\!\!\cdot\!\!\Bigg[\lambda^{2}+\frac{a}{2}\Big(\frac{3\alpha}{\beta a r^{2}}\!-\!1\Big)\!\Bigg] \;\;\; - \;\;\nonumber \\
&4a\Big(1\!-\!\frac{\alpha}{\beta a r^{2}}\Big)\!\!\cdot\!\!\Bigg[\lambda-\!\frac{i\sqrt{a}}{4}\Bigg]\!\!\cdot\!\!\Bigg[\frac{i\sqrt{a}}{4}-\!\alpha-\!\lambda\Bigg]=0.                         
\end{align}
Note that the complex-conjugate of $\lambda$ is also a solution of Eqs.(\ref{eq:LinRing1}-\ref{eq:LinRing2}). 

In general, the milling state on the sphere is linearly stable if there are no solutions to Eq.(\ref{eq:PeriodicMode}) with $Re[\lambda]\!>\!0$. By dividing $\lambda$ by $\alpha$ in Eq.(\ref{eq:PeriodicMode}), we can see that the spectrum for the spatially periodic mode is effectively a function of two positive parameters only, $\mu\equiv a/\alpha^{2}$ and $\nu \equiv \alpha/(\beta a r^{2})$, i.e. $\lambda/\alpha\!=\!L(\mu,\nu)$ and has four solutions. Moreover, in the relevant parameter region $\nu < 1$ 
the only change in stability occurs through a Hopf bifurcation\cite{Kuznetsov1,StrogatzBook,WigginsBook}. We return to what happens when $\nu = 1$ 
in Sec.\ref{sec:single}. 

Generically, a Hopf bifurcation occurs when $\lambda\!=\!\pm i\omega$, with $\omega\!\neq\!0$. It is easy to verify that this condition\cite{Note2} is satisfied in Eq.(\ref{eq:PeriodicMode}) when  
\begin{align}
\label{eq:Hopf}
\nu_H =  \frac{3}{8}
\end{align}
Importantly, the Hopf bifurcation uncovered is a curvature-induced instability; it does not occur in flat space. For instance, if one takes $r\!\rightarrow\!\infty$, Eq.(\ref{eq:Hopf}) can only be satisfied in the trivial infinite-speed or zero-coupling limits. Also, it is worth mentioning that since Eqs.(\ref{eq:LinRing1}-\ref{eq:LinRing2}) describe the linearized dynamics in co-moving reference frames, the Hopf bifurcations entailed by Eq.(\ref{eq:Hopf}) are generalized Hopf bifurcations of LCs (leading to motion on a high-dimensional torus) . However, for brevity and to avoid double usage of ``torus", we use the short-hand, Hopf, throughout. 

\begin{figure}[h]
\centering
\includegraphics[scale=0.26]{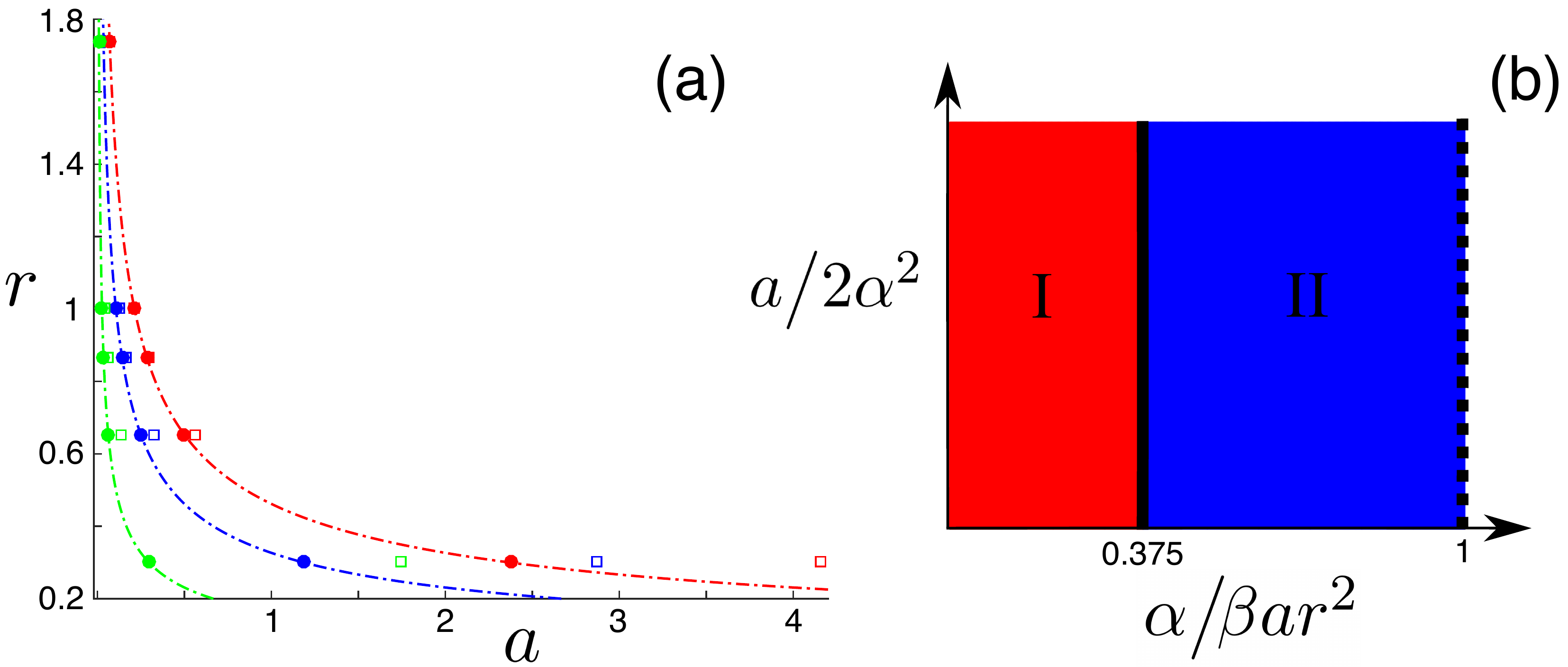}
\caption{(Color online) Milling stability on the sphere. (a) Hopf bifurcation curves are drawn with dashed lines for $\alpha\!=\!0.05$ (green, bottom), $\alpha\!=\!0.20$  (blue, middle), and $\alpha\!=\!0.40$  (red, top). In each case $\beta\!=\!5.0$.  Points denote simulation-determined stability changes for $N\!=\!600$: milling with all agents rotating in the same direction (circles), and milling with half the agents rotating in the opposite direction (squares). (b) Stability diagram. Region (I, red) has no unstable modes, and region (II, blue) has two unstable modes. (I) and (II) are separated by the Hopf bifurcation, Eq.(\ref{eq:Hopf}), drawn with a solid black line. The milling state exists to the left of the dashed-line.}
\label{fig:SphStab}
\end{figure}

Example Hopf curves are shown in Fig.\ref{fig:SphStab}(a) with dashed lines for several values of $\alpha$. For each value of $\alpha$, the milling state is predicted to be stable for all larger values of $a$ (i.e., moving to the right of the dashed-lines at fixed $r$). The circles denote simulation-determined transition points: the smallest $a(r)$ for which a swarm of 600 agents, initially prepared in the milling state with a small random perturbation (i.e., independent and uniformly distributed $A_{j}$ and $B_{j}$ over $[-10^{-5},10^{-5}]$), returns to a milling state after an integration time of $t=10000$. Predictions from Eq.(\ref{eq:Hopf}) show excellent agreement with simulation results. Similarly determined transition points for a milling state in which half the agents rotate in one direction, and half rotate in the opposite direction, are shown with squares. We can see that the Hopf-transition line still gives a good approximation for this more general case, especially for larger values of $r$. 

\begin{figure}[h]
\centering
\includegraphics[scale=0.30]{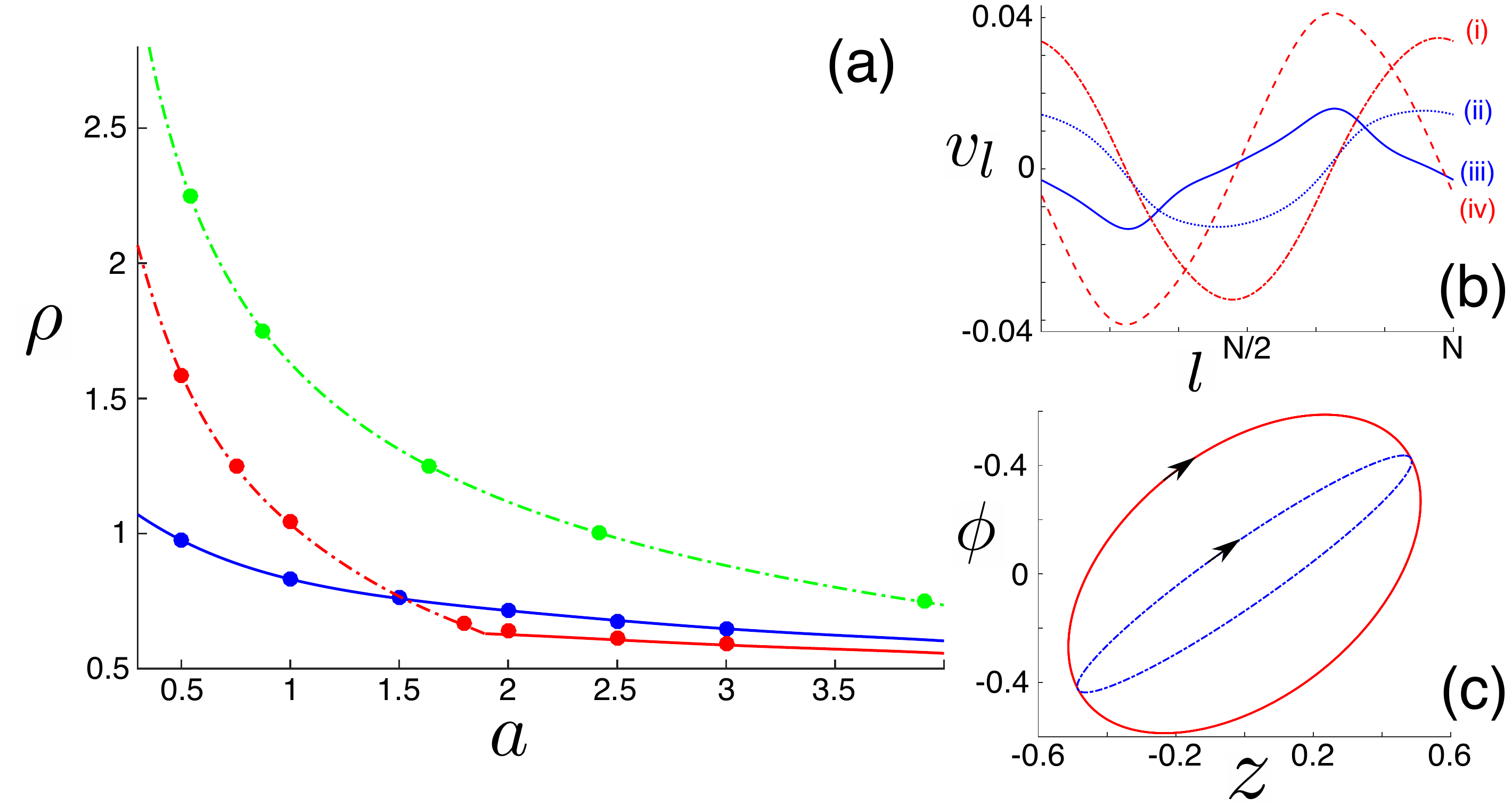}
\caption{(Color online) Milling stability on the cylinder. (a) Hopf bifurcation curves are drawn with dashed lines for $\alpha\!=\!4.0$ (green, top) and $\alpha\!=\!2.0$ (red, lower). SNpo curves are drawn with solid lines for $\alpha\!=\!0.5$ (blue, top) and $\alpha\!=\!2.0$ (red, lower). In each case $\beta\!=\!1.0$. Points denote simulation-determined stability changes for $N\!=\!600$, and follow the same color scheme. (b) Spatially periodic mode that changes stability in the Hopf bifurcation. Plotted are the mode components versus the agent number, $l$ when $a\!=\!0.34$, $\rho\!=\!1.94$, $\alpha\!=\!2$, and $\beta\!=\!1$: $Re[z_{l}]$ (solid blue, iii), $Im[z_{l}]$ (dotted blue, ii), $Re[\phi_{l}]$ (dashed red, iv), and $Im[\phi_{l}]$ (dotted-dashed red, i). (c) Stable (solid red, outer) and unstable (dashed blue, inner) limit cycles that collide in the saddle-node-of-periodic-orbits bifurcation. Parameters are $a\!=\!2$, $\rho\!=\!0.8$, $\alpha\!=\!0.5$, and $\beta\!=\!1$.}
\label{fig:CylStab}
\end{figure}

So far we have demonstrated that the spatially periodic modes that determine milling stability on the sphere are simple plane waves. On other surfaces the modes are periodic but not plane waves. 
Still, Hopf bifurcations occur, when such modes have complex-valued Floquet multipliers (FMs) that cross the unit circle. Floquet multipliers are the LC-analogs of the eigenspectrum for small variation around fixed-points in dynamical systems\cite{Kuznetsov1,StrogatzBook,Moore2005FloquetTA,WigginsBook}. Unlike the sphere, however, the FMs and their Hopf-points have to be determined numerically. In fact, for the Hopf bifurcations on the cylinder and torus we must compute the FMs for the whole multi-agent system; See App.\ref{sec:FM} for further details. 
 Examples for the cylinder are shown in Fig.\ref{fig:CylStab}. In panel (a), we plot the stability diagram for several values of $\alpha$. The dashed-lines are numerically determined Hopf bifurcations. For each value of $\alpha$, the milling state is predicted to be stable for all larger values of $a$ (i.e., moving to the right of the dashed-lines at fixed $\rho$). The circles are simulation-determined transition points-- computed in the same manner as for the sphere. As with the sphere, the LC linear-stability analysis agrees well with simulations. An example periodic mode that changes stability in a Hopf bifurcation is shown in panel (b). The mode is an eigenvector corresponding to one of the two FMs that crosses the unit circle. Plotted are the real and imaginary parts of the vector's $z_{l}$ and $\phi_{l}$-components. Similar results are shown for Hopf bifurcations on the torus in Sec.\ref{sec:torus}.
 
\subsection{\label{sec:single} Repeated-eigenvalue modes}
We have seen that the Hopf bifurcation of spatially periodic modes is qualitatively the same for milling states on different surfaces. A natural question concerns whether other types of stability changes occur, not due to a single complex pair of FMs, but the rest of the spectrum. To make progress, it is again useful to consider the sphere and the remaining solutions of Eqs.(\ref{eq:LinRing1}-\ref{eq:LinRing2}), first. After which, we can take a closer look at the cylinder, where a different bifurcation is possible. 

As mentioned previously, higher Fourier-modes have sums in Eqs.(\ref{eq:LinRing1}-\ref{eq:LinRing2}) that vanish. Consequently, Eqs.(\ref{eq:LinRing1}-\ref{eq:LinRing2}) have $N\!-\!1$ solutions with 
a common eigenvalue $s$, satisfying
\begin{align}
\label{eq:Repeated}
&\frac{s}{\alpha}\Bigg[\!\Big(\frac{s}{\alpha}\Big)^{\!2} + \frac{a}{\alpha^{2}}\frac{\alpha}{\beta a r^{2}}\Bigg]\Bigg[\!\Big(\frac{s}{\alpha}\Big)+2\Bigg] \;\;\; + \;\;\nonumber \\
&\frac{s}{\alpha}\frac{4a}{\alpha^{2}}\Bigg[1-\!\frac{\alpha}{\beta a r^{2}}\!\Bigg]\Bigg[\!\Big(\frac{s}{\alpha}\Big)+1\Bigg]=0.                         
\end{align}
Just as with Eq.(\ref{eq:PeriodicMode}), there are two effective parameters and four solutions to Eq.(\ref{eq:Repeated})-- giving a total of $4(N\!-\!1)$ such solutions, which fill out the remaining FM spectrum. One can check that all solutions have $Re[s]\!\leq\!0$ when $\nu\!=\!\alpha/[\beta a r^{2}]\!<\!1$. 

When $\nu \!=\!\alpha/(\beta a r^{2})\!=\!1$, a {\it degenerate} Hopf bifurcation occurs -- degenerate, since $2(N\!-\!1)$ FMs cross the unit circle {\it simultaneously}. Putting the whole spectrum of the milling state on the sphere together, we can now draw the complete stability diagram, shown in Fig.\ref{fig:SphStab}(b). The red region (I) has no unstable modes, and the blue region (II) has two unstable modes. The Hopf bifurcation for the periodic mode is drawn with a solid black line, while the degenerate Hopf bifurcation of the repeated part of the spectrum is drawn with a dashed black line. It is clear from Fig.\ref{fig:SphStab}(b), that the milling state on the sphere changes stability {\it only} through the Hopf bifurcation of spatially-periodic modes. Accounting for the rest of the spectrum does not change the stability picture.  

However, this is not true in general for milling on other surfaces. In particular, on the cylinder we find that ~${N\!-\!1}$ {\it real} FMs can approach unity in a degenerate SNpo bifurcation\cite{Kuznetsov1,StrogatzBook,WigginsBook}. An example FM-spectrum near bifurcation is shown in Sec.\ref{sec:FM} for reference. 

To track the SNpo numerically, it is sufficient to compute the FMs for a LC of the single-particle system, Eqs(\ref{eq:LCcyc1}-\ref{eq:LCcyc3}) -- making the computation significantly faster than for the Hopf bifurcation of spatially-periodic modes. Examples of SNpo curves are shown in Fig.\ref{fig:CylStab}(a) with solid lines. Simulation-determined transitions points are plotted with colored circles. As with the Hopf bifurcations, the milling state is predicted to be stable for all larger values of $a$ (i.e., moving to the right of the solid lines at fixed $\rho$). 

\subsubsection{\label{sec:SNpo} Stable and unstable cycle pairs}
An important hallmark of SNpo bifurcations is the collision and annihilation of two periodic orbits\cite{Kuznetsov1,StrogatzBook,WigginsBook}. As a consequence, we expect a stable milling-state LC to coalesce with an unstable LC, for swarms on the cylinder, both of which have $N\!-\!1$ FMs that approach unity from below and above, respectively, at bifurcation. Figure \ref{fig:CylStab}(c) plots an example of the stable and unstable LCs computed from Eqs.(\ref{eq:LCcyc1}-\ref{eq:LCcyc3}) using MATLAB's version of AUTO continuation software. It is important to note that the unstable LCs are distinct from the reflected LCs discussed in Sec.\ref{sec:Cycle}. Namely, on the cylinder there are two stable and two unstable milling states in general. 

Further quantitative insight can be obtained on the LC pairs by finding approximate, periodic solutions to Eqs(\ref{eq:LCcyc1}-\ref{eq:LCcyc3}). As mentioned previously, on surfaces LCs are generally not simple sinusoids. However, we can solve for the the first-harmonic of the Fourier series, $z(t)^{(LC)}\!\approx\!Z\sin(\omega t+\gamma)$ and $\phi(t)^{(LC)}\!\approx\!(Z/\rho)\sin\omega t$, with the assumption that higher harmonics make only small contributions. Substituting our ansatz into Eqs(\ref{eq:LCcyc1}-\ref{eq:LCcyc3}), multiplying the equations by $\sin\omega t$ (and $\cos\omega t$ respectively), and integrating over a period, we obtain the following three fixed-point equations:
\begin{align}
\label{eq:CylinderFPs1}
\omega^{2}-a=& \frac{1}{4}\beta Z^{2}\omega^{3}\sin2\gamma,\\
\label{eq:CylinderFPs2}
Z^{2}\omega^{2}=&\frac{4\alpha}{\beta (5+\cos2\gamma)},\\
\label{eq:CylinderFPs3}
\frac{Z\omega^{2}}{2\rho a}\!\Big[1\!+\!\frac{\omega^{2}-a}{\omega^{2}}\Big]=&\int_{0}^{2\pi}\!\!\!\!\!\cos\Big(\!\frac{Z}{\rho}\sin\sigma'\!\Big)\frac{d\sigma'}{2\pi}\;\;\cdot\nonumber\\                      
&\int_{0}^{2\pi}\!\!\!\!\!\sin\Big(\!\frac{Z}{\rho}\sin\sigma\!\Big)\frac{\sin \sigma d\sigma}{2\pi}, 
\end{align}
which can be solved for the approximate parameters of both stable and unstable milling states on the cylinder. 

\begin{figure}[h]
\centering
\includegraphics[scale=0.28]{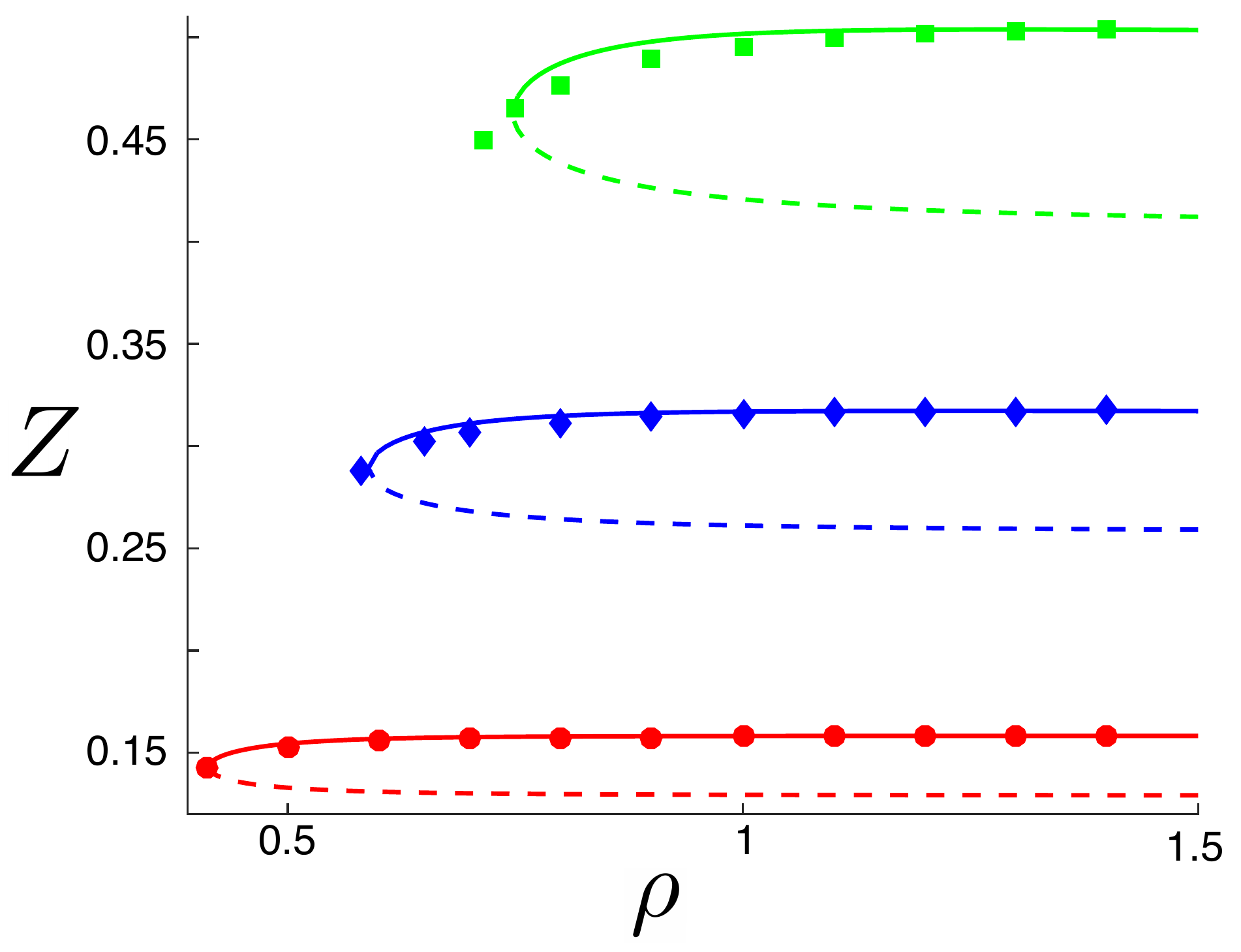}
\caption{(Color online) Saddle-node-of-periodic-orbits diagram for milling states on the cylinder. Predictions for the limit-cycle amplitudes, Eqs.(\ref{eq:CylinderFPs1}-\ref{eq:CylinderFPs3}), are shown for the stable (solid) and unstable (dashed) cycles with $a\!=\!20$ (red, lower), $a\!=\!5$ (blue, middle), and $a\!=\!2$ (green, top). Points denote simulation amplitudes for a swarm of $N\!=\!600$ agents. Parameters are $\alpha\!=\!0.5$ and $\beta\!=\!1.0$.}
\label{fig:FirstAmpl}
\end{figure}

Equations (\ref{eq:CylinderFPs1}-\ref{eq:CylinderFPs3}) have two non-trivial solutions that meet at a critical curvature, corresponding to the SNpo bifurcation, as expected. 
Example solutions are shown in Fig.\ref{fig:FirstAmpl} for large (red), medium (blue), and small (green) coupling. The stable amplitude is shown with a solid line and the unstable with a dashed line. The agreement between the first-harmonic approximation and large-agent simulations (points) is fair in general, and increases with the coupling.   

It is clear from Fig.\ref{fig:FirstAmpl} that as $\rho$ is increased above the SNpo bifurcation, the amplitudes converge quickly to their asymptotic values. The limiting solutions correspond to $\gamma\!\rightarrow 0,$ $\pi/2$: 
\begin{align}
\label{eq:Limiting}
Z^{2}=\frac{2\alpha}{3\beta a}\;\; \text{,} \;\; \frac{\alpha}{\beta a}. 
\end{align}
as $\rho\rightarrow\infty$. The asymptotic results imply an important property of milling on the cylinder: unstable states exist even when mean curvature is {\it weak}. The persistence of such dynamically unstable orbits makes swarming on cylindrical surfaces qualitatively different from, e.g., the sphere, where no such orbits exist.          
   
\section{\label{sec:Conclusion} CONCLUSION}
In this work we studied the stability of stationary, milling patterns in self-propelled swarms constrained to move on surfaces. Using Lagrangian mechanics, we found that with simple attractive forces between agents such patterns correspond to surface-dependent limit cycles where agents in a swarm are splayed along different points on the same cycle. We showed that the constraint of curvature can destabilize milling swarms, and does so in both generic and geometry-specific ways. In the former, a spatially periodic mode of the milling state becomes unstable in a generalized Hopf bifurcation. This bifurcation was demonstrated and analyzed for swarming on the sphere, cylinder, and torus, as examples. On the other hand, we found that on cylindrical surfaces unstable milling patterns exist, even in the limit of small curvature, and can merge with stable milling states in a saddle-node-of-periodic-orbits bifurcation. Our analytical and numerical results were verified in detail with large multi-agent simulations.  

To our knowledge, these are the first formal bifurcation results for swarming patterns in general geometries, and they hint at a more general theory for swarming stability on Riemannian manifolds, especially in the limit of weak mean-curvature. Extending the Lagrangian formulation of the swarm dynamics in terms of geodesic potentials on arbitrary surfaces, as well as a geodesically defined center-of-mass that is constrained to the surface, will allow for a more accurate analysis of milling patterns on surfaces with higher sectional curvature variability. We expect our methods to be useful for analyzing patterns in other swarming problems where both the patterns themselves and the constraints imposed are expressible in terms of symmetries and generalized coordinates. For instance, an important future generalization of the swarms studied here is the inclusion of repulsive forces, since such forces are known to create a variety of milling patterns on surfaces beyond the simple splayed limit cycles that we analyzed. 

Considering the patterns of swarm dynamics on specific classes of surfaces used in practical applications is another possible direction of future research. For instance, piece-wise polynomial surfaces, such as splines, or more generally rational splines can be of interest for ground robotics applications. Parametrically controlling swarms of mobile robots to cooperatively move over complex surfaces and terrains in the future will require studying the effects of communication topology, both dynamic and incomplete, as well as uncertainty and noise. Of course, many open questions remain. Yet, our approach shows that general stability analysis for large self-propelled swarms on surfaces is possible and we demonstrate how it might be done.  

JH and IBS were supported by the U.S. Naval Research Laboratory funding (N0001419WX00055) and the Office of Naval Research (N0001419WX01166), (N0001419WX01322), and (N0001420WX01237). TE was supported through the U.S Naval Research Laboratory Karles Fellowship. SK was supported through the GMU Provost PhD award as part of the Industrial Immersion Program. GS was supported through the Office of Naval Research funding (N0001420WX01237).\\

\section{\label{sec:App}APPENDIX}
\subsection{\label{sec:torus} Swarming on the torus}
As defined in Sec.\ref{sec:Cycle}, the torus has two radial parameters: $b$ and $c$. The former is the tube radius, while the latter is the radius from the center of the hole to the center of the tube.   
Using the coordinate transformation, $(x_{l},y_{l},z_{l})_{t}\!=\!\left((c+b\cos\theta_{l})\!\cos\phi_{l},(c+b\cos\theta_{l})\!\sin\phi_{l},b\sin\theta_{l}\right)$, where $\theta$ is an azimuthal angle and $\phi$ is the angle inside of the tube, we find the swarming Eqs.(\ref{eq:GeneralLagrange}-\ref{eq:Lagrangian}): 
\begin{align}
\label{eq:Torus1}
&\ddot{\phi}_{l}= \dot{\phi}_{l}\Bigg[\alpha-\!\beta\Big(b^{2}\dot{\theta}_{l}^{2}+(c+\!b\cos\theta_{l})^{2}\dot{\phi}_{l}^{2}\Big)\!\Bigg]\;\;\;+ \nonumber \\
&\frac{a}{N(c\!+\!b\cos\theta_{l})}\!\sum_{j}(c+\!b\cos\theta_{j})\!\sin(\phi_{j}\!-\!\phi_{l})\;\;\;+\nonumber \\ 
&\frac{2b\sin\theta_{l}}{(c+b\cos\theta_{l})}\dot{\phi_{l}}\dot{\theta_{l}},\\
\label{eq:Torus2}
&\ddot{\theta}_{l}= \dot{\theta}_{l}\Bigg[\alpha-\!\beta\Big(b^{2}\dot{\theta}_{l}^{2}+(c+\!b\cos\theta_{l})^{2}\dot{\phi}_{l}^{2}\Big)\!\Bigg] \;\;\;+\nonumber\\ 
&\frac{a}{N}\!\sum_{j}\!\sin(\theta_{j}\!-\!\theta_{l})\;\;-\;\;\frac{(c+b\cos\theta_{l})\sin\theta_{l}}{b}\dot{\phi_{l}}^{2}\;\;\;+\nonumber\\
&\frac{a\sin\theta_{l}}{Nb}\!\sum_{j}(1\!-\!\cos(\phi_{j}\!-\!\phi_{l}))(c+\!b\cos\theta_{j}).
\end{align}

Furthermore the single-particle system, used to compute LCs associated with milling, can be found by substituting Eq.(\ref{eq:SelfCons}) into Eqs.(\ref{eq:Torus1}-\ref{eq:Torus2}). By choosing a solution with $\left<z\right>\!=\!\left<y\right>\!=\!0$, the single-particle system only depends on the time-average of $x$, $\left<x\right>\!=\!\left<(c+b\cos\theta)\!\cos\phi\right>$, or 
\begin{align}
\label{eq:SPTorus1}
&\ddot{\phi}= \dot{\phi}\Bigg[\alpha-\!\beta\Big(b^{2}\dot{\theta}^{2}+(c+\!b\cos\theta)^{2}\dot{\phi}^{2}\Big)\!\Bigg]\;\;\;+ \nonumber \\
&\frac{2b\dot{\phi}\;\dot{\theta}\sin\theta-a\!\left<x\right>\sin\phi}{c+b\cos\theta}\;,\\
\label{eq:SPTorus2}
&\ddot{\theta}= \dot{\theta}\Bigg[\alpha-\!\beta\Big(b^{2}\dot{\theta}^{2}+(c+\!b\cos\theta)^{2}\dot{\phi}^{2}\Big)\!\Bigg] \;\;\;+\nonumber\\
&\frac{a}{b}\big(c-\!\left<x\right>\!\cos\phi\big)\sin\theta-\frac{(c+b\cos\theta)\sin\theta}{b}\dot{\phi}^{2}.
\end{align}
Equations (\ref{eq:SPTorus1}-\ref{eq:SPTorus2}) were used for the theory comparisons in Fig.\ref{fig:Patterns}(c). 

\begin{figure}[h]
\centering
\includegraphics[scale=0.265]{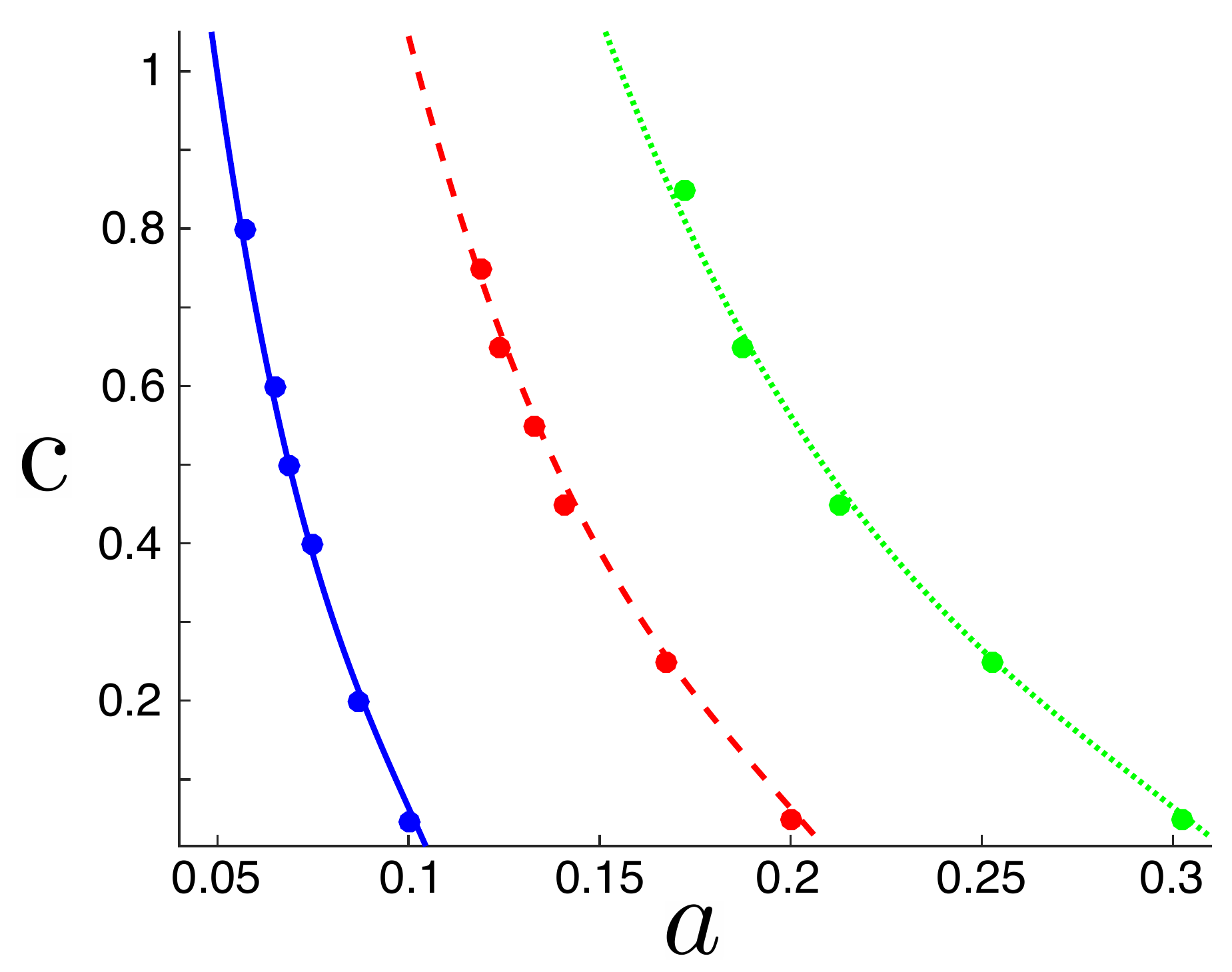}
\caption{(Color online) Milling stability on the torus. Hopf bifurcation curves are drawn with solid lines for $\alpha\!=\!0.2$ (blue, left), dashed lines for $\alpha\!=\!0.4$  (red, middle), and dotted lines for $\alpha\!=\!0.6$  (green, right). In each case $b\!=\!\!1.0$ and $\beta\!=\!5.0$. Points denote simulation-determined stability changes for $N\!=\!600$, and follow the same color scheme.}
\label{fig:Torus}
\end{figure}

Note that if we take $c\!\rightarrow\!0$ and $\theta_{j}\!\rightarrow\!\theta_{j}-\pi/2 \;\forall j$, in Eqs.(\ref{eq:Torus1}-\ref{eq:Torus2}), we get the spherical system, Eq.(\ref{eq:Sph1}-\ref{eq:Sph2}), back with $r\!=\!b$.  Consequently, for relatively small $c$ we expect to qualitatively reproduce the stability picture for milling states on the sphere. Namely, such states will change stability through Hopf bifurcations of spatially-periodic modes, as discussed in Sec.\ref{sec:periodic} for the sphere and cylinder. Three examples of Hopf-curves that demonstrate this prediction are shown in Fig.\ref{fig:Torus}, and agree well with large multi-agent simulations.  

\subsection{\label{sec:Uncoupled}Uncoupled dynamics}
In the uncoupled limit ($a\!=\!0$) Eq.(\ref{eq:GeneralLagrange}) becomes: 
\begin{align}
\label{eq:Uncoupled}
\frac{d}{dt}\Bigg[v\frac{\partial v}{\partial \dot{q}^{k}}\Bigg]=v\frac{\partial v}{\partial \dot{q}^{k}}\Big(\alpha-\beta v^{2}\Big).         
\end{align}
Note that we have dropped the agent subscript, $l$. From generic initial conditions, we find that the forces on the right-hand-side of Eq.(\ref{eq:Uncoupled}) vanish as $t\!\rightarrow\!\infty$, resulting in a constant speed $v\!=\!\sqrt{\alpha/\beta}$. By Hamilton's principle, we know that $\frac{d}{dt}\big[v\frac{\partial v}{\partial \dot{q}^{k}}\big]=0$, describes trajectories that {\it extremize} the integral $S[\bold{q},\bold{\dot{q}}]\!=\!\frac{1}{2}\!\int\!v^{2}dt$, which is the action for a free-particle with speed $v$. Since $v\!=\!\sqrt{\alpha/\beta}=\frac{ds}{dt}$, where $ds$ is the differential arc length along a surface,   
\begin{align}
\label{eq:Action}
S=\frac{1}{2}\sqrt{\frac{\alpha}{\beta}}\int \!ds.                       
\end{align}
Equation (\ref{eq:Action}) implies that, in the absence of interactions and in the long-time limit, the integral which is extremized along an agent's trajectory is proportional to the arc length. Hence, such trajectories are surface geodesics.  

\subsection{\label{sec:FM} Numerical Floquet multipliers}
Computing FMs numerically amounts to solving an eigenvalue problem for a linear, discrete-dynamical system. We start by considering the repeated part of the Floquet spectrum for milling states, since we can restrict ourselves to the single-particle systems, e.g., Eqs.(\ref{eq:Cyl1}-\ref{eq:Cyl2}) and Eqs.(\ref{eq:SPTorus1}-\ref{eq:SPTorus2}). Recall that this part of the spectrum is relevant for the SNpo bifurcations discussed in Sec.\ref{sec:single}. Let us define a 4-dimensional phase-space vector (coordinates plus velocities) for the single-particle system, $\bold{P}\!=\!\left(q^{1},\dot{q}^{1},q^{2},\dot{q}^{2}\right)$. Note that the superscript is a label for the generalized coordinates (not an exponent), and we drop the agent subscript $l$. Our goal is to compute the characteristic dynamics for small variations around LCs, $\boldsymbol{\epsilon}(t)\!=\!\bold{P}^{(LC)}(t;\bold{R})\!-\!\bold{P}(t)$, where $(LC)$ denotes the phase-space coordinates evaluated on a LC. Since the dynamics are assumed to occur near LCs, for sufficiently small $\boldsymbol{\epsilon}(t)$, we can employ Floquet's theorem at lowest order in $\boldsymbol{\epsilon}(t)$, which states that $\boldsymbol{\epsilon}(t+T)\!=\!\underline{\bold{M}}\boldsymbol{\epsilon}(t)$ \cite{Kuznetsov1,StrogatzBook,Moore2005FloquetTA,WigginsBook}, where $\underline{\bold{M}}$ is a constant, non-singular monodromy matrix of dimension $4$x$4$ and $T$ is the period of the LC.  

The first step is to compute the LC as described in Sec.\ref{sec:Cycle} for given parameters. The second step is to compute $\underline{\bold{M}}$. The repeated FMs of the milling state are simply the eigenvalues of $\underline{\bold{M}}$.   
A straightforward way to determine $\underline{\bold{M}}$ is to compute four perturbations, $\boldsymbol{\epsilon}_{1}(T)$, $\boldsymbol{\epsilon}_{2}(T)$, $\boldsymbol{\epsilon}_{3}(T)$, and $\boldsymbol{\epsilon}_{4}(T)$. Each can be found by  {\it integrating the single-particle system} over one period from the initial conditions $\boldsymbol{\epsilon}_{1}(0)\!=\epsilon\left(1,0,0,0\right)$, $\boldsymbol{\epsilon}_{2}(0)\!=\epsilon\left(0,1,0,0\right)$, $\boldsymbol{\epsilon}_{3}(0)\!=\epsilon\left(0,0,1,0\right)$, and $\boldsymbol{\epsilon}_{4}(0)\!=\epsilon\left(0,0,0,1\right)$, respectively, for some small $\epsilon >0$. The integrated perturbations fill out the columns of $\underline{\bold{M}}$; in particular $\underline{\bold{M}}_{jn}=[\boldsymbol{\epsilon}_{n}(T)]_{j}/\epsilon$, where $j, n \in\{1,2,3,4\}$. For reference, $\epsilon\!=\!10^{-4}$ in all of the computations in this work. Lastly, to compute SNPo bifurcation-points on the cylinder for fixed $\rho$, we simply reduce $a$ in small steps from some starting value until the second largest eigenvalue of $\underline{\bold{M}}$ is equal to $1$ (within some small error tolerance).
\begin{figure}[t]
\vspace{0.45cm}
\centering
\includegraphics[scale=0.239]{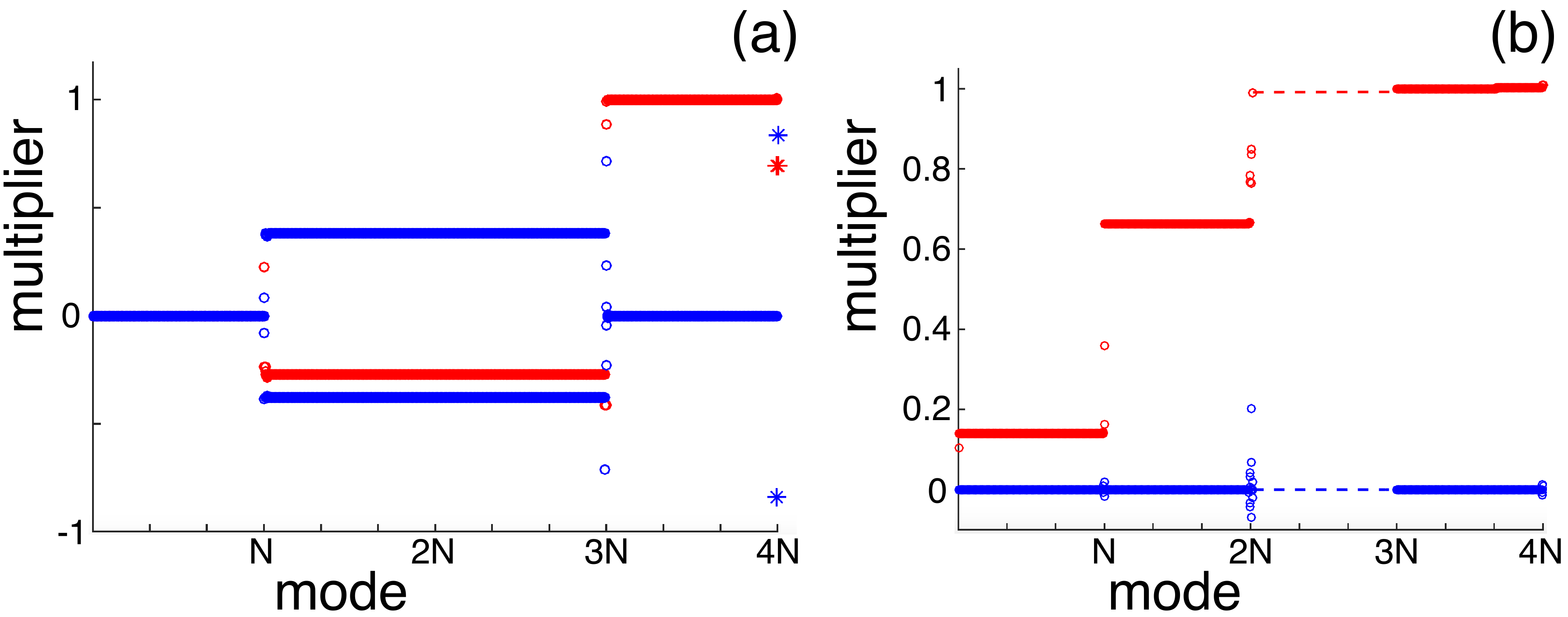}
\caption{(Color online) Example Floquet multipliers for milling states on the cylinder, the full multi-agent system, Eqs.(\ref{eq:Cyl1}-\ref{eq:Cyl2}). (a) Multipliers near a Hopf bifurcation, $a\!=\!0.34$, $\rho\!=\!1.94$, $\alpha\!=\!2$, and $\beta\!=\!1$. The real and imaginary parts of the multipliers are plotted with red and blue circles, respectively. The two multipliers that cross the unit circle at bifurcation are plotted with asterisks. (b) Multipliers near a SNpo bifurcation, $a\!=\!3$, $\rho\!=\!0.645$, $\alpha\!=\!0.5$, and $\beta\!=\!1$. The repeated (real) multipliers that approach $1$ at bifurcation are plotted with a dashed line.}
\label{fig:FMs}
\end{figure} 

On the other hand, for the Hopf bifurcations we must compute the FMs for the full multi-agent LC, since the spatially-periodic modes that change stability entail collective oscillations of all agents. Keep in mind that the repeated part of the full spectrum will equal the single-particle FMs discussed in the previous paragraph for large $N$. For reference, $N\!=\!300$ in all FM-computations in this work, which is large enough to produce negligibly small finite-size effects. Our approach is essentially the same as for the single-particle system, except that monodromy matrix in this case is $4N$x$4N$ dimensional. Following the same procedure, let us define the full phase-space vector $\bold{X}\!=\!\left(q_{1}^{1},\dot{q}_{1}^{1},q_{1}^{2},\dot{q}_{1}^{2},q_{2}^{1},\dot{q}_{2}^{1},q_{2}^{2},\dot{q}_{2}^{2},...,q_{N}^{1},\dot{q}_{N}^{1},q_{N}^{2},\dot{q}_{N}^{2}\right)$. The first step is to construct the complete milling state from the computed single-particle LC. The assumption of Eq.(\ref{eq:SelfCons}) implies that agents are splayed uniformly in time along a LC, and therefore we take
\begin{align}
\bold{X}(t;\bold{R})^{(LC)}=\big(&\bold{P}^{(LC)}(t+[1\!-\!1]T/N;\bold{R}),\nonumber\\
&\bold{P}^{(LC)}(t+[2\!-\!1]T/N;\bold{R}),...,\nonumber\\ 
&\bold{P}^{(LC)}(t+[N\!-\!1]T/N;\bold{R})\big).
\end{align}
Next, we define the perturbation $\boldsymbol{\delta}(t)\!=\!\bold{X}(t;\bold{R})^{(LC)}\!-\!\bold{X}(t)$, which has dynamics $\boldsymbol{\delta}(t+T)\!=\!\underline{\bold{D}}\boldsymbol{\delta}(t)$, for small $\boldsymbol{\delta}$ by Floquet's theorem. The matrix $\underline{\bold{D}}$ can be determined by computing $4N$ integrated perturbations, $\boldsymbol{\delta}_{l}(T)$, where $\boldsymbol{\delta}_{l}(t\!=\!0)\!=\!\epsilon{\left(0,0,...,1_{\;l},0,...\right)}$ with $l\in\{1,2,...,4N\}$. In contrast to the preceding paragraph, these perturbations are integrated in the full system, Eqs.(\ref{eq:GeneralLagrange}-\ref{eq:Lagrangian}). The integrated perturbations fill out the columns of $\underline{\bold{D}}$, such that  $\underline{\bold{D}}_{jl}=[\boldsymbol{\delta}_{l}(T)]_{j}/\epsilon$, where $j\in\{1,2,...,4N\}$. As before, the FMs for the full system are the eigenvalues of $\underline{\bold{D}}$.    

Finally, to compute Hopf-bifurcation points on the cylinder or torus for fixed curvature, we simply reduce $a$ in small steps from some starting value until a pair of eigenvalues of $\underline{\bold{D}}$, not in the repeated part of the spectrum, have unit absolute value (within some small error tolerance). Examples of the FMs for both milling-state bifurcations on the cylinder are shown in Fig.\ref{fig:FMs} for the full multi-agent system. As mentioned above, the $N$ real FMs that approach $1$ in panel (b), are identical to the single-particle FMs.
          
\bibliography{millingSwarmsOnSurfaces,DelayBistabilityPRE1}

\begin{thebibliography}{60}%
\makeatletter
\providecommand \@ifxundefined [1]{%
 \@ifx{#1\undefined}
}%
\providecommand \@ifnum [1]{%
 \ifnum #1\expandafter \@firstoftwo
 \else \expandafter \@secondoftwo
 \fi
}%
\providecommand \@ifx [1]{%
 \ifx #1\expandafter \@firstoftwo
 \else \expandafter \@secondoftwo
 \fi
}%
\providecommand \natexlab [1]{#1}%
\providecommand \enquote  [1]{``#1''}%
\providecommand \bibnamefont  [1]{#1}%
\providecommand \bibfnamefont [1]{#1}%
\providecommand \citenamefont [1]{#1}%
\providecommand \href@noop [0]{\@secondoftwo}%
\providecommand \href [0]{\begingroup \@sanitize@url \@href}%
\providecommand \@href[1]{\@@startlink{#1}\@@href}%
\providecommand \@@href[1]{\endgroup#1\@@endlink}%
\providecommand \@sanitize@url [0]{\catcode `\\12\catcode `\$12\catcode
  `\&12\catcode `\#12\catcode `\^12\catcode `\_12\catcode `\%12\relax}%
\providecommand \@@startlink[1]{}%
\providecommand \@@endlink[0]{}%
\providecommand \url  [0]{\begingroup\@sanitize@url \@url }%
\providecommand \@url [1]{\endgroup\@href {#1}{\urlprefix }}%
\providecommand \urlprefix  [0]{URL }%
\providecommand \Eprint [0]{\href }%
\providecommand \doibase [0]{http://dx.doi.org/}%
\providecommand \selectlanguage [0]{\@gobble}%
\providecommand \bibinfo  [0]{\@secondoftwo}%
\providecommand \bibfield  [0]{\@secondoftwo}%
\providecommand \translation [1]{[#1]}%
\providecommand \BibitemOpen [0]{}%
\providecommand \bibitemStop [0]{}%
\providecommand \bibitemNoStop [0]{.\EOS\space}%
\providecommand \EOS [0]{\spacefactor3000\relax}%
\providecommand \BibitemShut  [1]{\csname bibitem#1\endcsname}%
\let\auto@bib@innerbib\@empty
\bibitem [{\citenamefont {Polezhaev}\ \emph {et~al.}(2006)\citenamefont
  {Polezhaev}, \citenamefont {Pashkov}, \citenamefont {Lobanov},\ and\
  \citenamefont {Petrov}}]{Polezhaev}%
  \BibitemOpen
  \bibfield  {author} {\bibinfo {author} {\bibfnamefont {A.}~\bibnamefont
  {Polezhaev}}, \bibinfo {author} {\bibfnamefont {R.}~\bibnamefont {Pashkov}},
  \bibinfo {author} {\bibfnamefont {A.~I.}\ \bibnamefont {Lobanov}}, \ and\
  \bibinfo {author} {\bibfnamefont {I.~B.}\ \bibnamefont {Petrov}},\
  }\href@noop {} {\bibfield  {journal} {\bibinfo  {journal} {Int. J. Dev.
  Bio.}\ }\textbf {\bibinfo {volume} {50}},\ \bibinfo {pages} {309} (\bibinfo
  {year} {2006})}\BibitemShut {NoStop}%
\bibitem [{\citenamefont {Li}\ and\ \citenamefont
  {Sayed}(2012)}]{Li_Sayed_2012}%
  \BibitemOpen
  \bibfield  {author} {\bibinfo {author} {\bibfnamefont {J.}~\bibnamefont
  {Li}}\ and\ \bibinfo {author} {\bibfnamefont {A.~H.}\ \bibnamefont {Sayed}},\
  }\href {\doibase 10.1186/1687-6180-2012-18} {\bibfield  {journal} {\bibinfo
  {journal} {EURASIP Journal on Advances in Signal Processing}\ }\textbf
  {\bibinfo {volume} {2012}},\ \bibinfo {pages} {18} (\bibinfo {year}
  {2012})}\BibitemShut {NoStop}%
\bibitem [{\citenamefont {Theraulaz}\ \emph {et~al.}(2002)\citenamefont
  {Theraulaz}, \citenamefont {Bonabeau}, \citenamefont {Nicolis}, \citenamefont
  {Sol{\'e}}, \citenamefont {Fourcassi{\'e}}, \citenamefont {Blanco},
  \citenamefont {Fournier}, \citenamefont {Joly}, \citenamefont
  {Fern{\'a}ndez}, \citenamefont {Grimal}, \citenamefont {Dalle},\ and\
  \citenamefont {Deneubourg}}]{Theraulaz2002}%
  \BibitemOpen
  \bibfield  {author} {\bibinfo {author} {\bibfnamefont {G.}~\bibnamefont
  {Theraulaz}}, \bibinfo {author} {\bibfnamefont {E.}~\bibnamefont {Bonabeau}},
  \bibinfo {author} {\bibfnamefont {S.~C.}\ \bibnamefont {Nicolis}}, \bibinfo
  {author} {\bibfnamefont {R.~V.}\ \bibnamefont {Sol{\'e}}}, \bibinfo {author}
  {\bibfnamefont {V.}~\bibnamefont {Fourcassi{\'e}}}, \bibinfo {author}
  {\bibfnamefont {S.}~\bibnamefont {Blanco}}, \bibinfo {author} {\bibfnamefont
  {R.}~\bibnamefont {Fournier}}, \bibinfo {author} {\bibfnamefont {J.-L.}\
  \bibnamefont {Joly}}, \bibinfo {author} {\bibfnamefont {P.}~\bibnamefont
  {Fern{\'a}ndez}}, \bibinfo {author} {\bibfnamefont {A.}~\bibnamefont
  {Grimal}}, \bibinfo {author} {\bibfnamefont {P.}~\bibnamefont {Dalle}}, \
  and\ \bibinfo {author} {\bibfnamefont {J.-L.}\ \bibnamefont {Deneubourg}},\
  }\href {\doibase 10.1073/pnas.152302199} {\bibfield  {journal} {\bibinfo
  {journal} {Proceedings of the National Academy of Sciences}\ }\textbf
  {\bibinfo {volume} {99}},\ \bibinfo {pages} {9645} (\bibinfo {year}
  {2002})},\ \Eprint
  {http://arxiv.org/abs/https://www.pnas.org/content/99/15/9645.full.pdf}
  {https://www.pnas.org/content/99/15/9645.full.pdf} \BibitemShut {NoStop}%
\bibitem [{\citenamefont {Topaz}\ \emph {et~al.}(2012)\citenamefont {Topaz},
  \citenamefont {D'Orsogna}, \citenamefont {Edelstein-Keshet},\ and\
  \citenamefont {Bernoff}}]{Topaz2012}%
  \BibitemOpen
  \bibfield  {author} {\bibinfo {author} {\bibfnamefont {C.~M.}\ \bibnamefont
  {Topaz}}, \bibinfo {author} {\bibfnamefont {M.~R.}\ \bibnamefont
  {D'Orsogna}}, \bibinfo {author} {\bibfnamefont {L.}~\bibnamefont
  {Edelstein-Keshet}}, \ and\ \bibinfo {author} {\bibfnamefont {A.~J.}\
  \bibnamefont {Bernoff}},\ }\href {\doibase 10.1371/journal.pcbi.1002642}
  {\bibfield  {journal} {\bibinfo  {journal} {PLoS Comput. Biol.}\ }\textbf
  {\bibinfo {volume} {8}},\ \bibinfo {pages} {1} (\bibinfo {year}
  {2012})}\BibitemShut {NoStop}%
\bibitem [{\citenamefont {Tunstrøm}\ \emph {et~al.}(2013)\citenamefont
  {Tunstrøm}, \citenamefont {Katz}, \citenamefont {Ioannou}, \citenamefont
  {Huepe}, \citenamefont {Lutz},\ and\ \citenamefont {Couzin}}]{Couzin2013}%
  \BibitemOpen
  \bibfield  {author} {\bibinfo {author} {\bibfnamefont {K.}~\bibnamefont
  {Tunstrøm}}, \bibinfo {author} {\bibfnamefont {Y.}~\bibnamefont {Katz}},
  \bibinfo {author} {\bibfnamefont {C.~C.}\ \bibnamefont {Ioannou}}, \bibinfo
  {author} {\bibfnamefont {C.}~\bibnamefont {Huepe}}, \bibinfo {author}
  {\bibfnamefont {M.~J.}\ \bibnamefont {Lutz}}, \ and\ \bibinfo {author}
  {\bibfnamefont {I.~D.}\ \bibnamefont {Couzin}},\ }\href {\doibase
  10.1371/journal.pcbi.1002915} {\bibfield  {journal} {\bibinfo  {journal}
  {PLOS Computational Biology}\ }\textbf {\bibinfo {volume} {9}},\ \bibinfo
  {pages} {1} (\bibinfo {year} {2013})}\BibitemShut {NoStop}%
\bibitem [{\citenamefont {Calovi}\ \emph {et~al.}(2014)\citenamefont {Calovi},
  \citenamefont {Lopez}, \citenamefont {Ngo}, \citenamefont {Sire},
  \citenamefont {Chat{\'{e}}},\ and\ \citenamefont {Theraulaz}}]{Calovi2014}%
  \BibitemOpen
  \bibfield  {author} {\bibinfo {author} {\bibfnamefont {D.~S.}\ \bibnamefont
  {Calovi}}, \bibinfo {author} {\bibfnamefont {U.}~\bibnamefont {Lopez}},
  \bibinfo {author} {\bibfnamefont {S.}~\bibnamefont {Ngo}}, \bibinfo {author}
  {\bibfnamefont {C.}~\bibnamefont {Sire}}, \bibinfo {author} {\bibfnamefont
  {H.}~\bibnamefont {Chat{\'{e}}}}, \ and\ \bibinfo {author} {\bibfnamefont
  {G.}~\bibnamefont {Theraulaz}},\ }\href {\doibase
  10.1088/1367-2630/16/1/015026} {\bibfield  {journal} {\bibinfo  {journal}
  {New Journal of Physics}\ }\textbf {\bibinfo {volume} {16}},\ \bibinfo
  {pages} {015026} (\bibinfo {year} {2014})}\BibitemShut {NoStop}%
\bibitem [{\citenamefont {Cavagna}\ \emph {et~al.}(2015)\citenamefont
  {Cavagna}, \citenamefont {Del~Castello}, \citenamefont {Giardina},
  \citenamefont {Grigera}, \citenamefont {Jelic}, \citenamefont {Melillo},
  \citenamefont {Mora}, \citenamefont {Parisi}, \citenamefont {Silvestri},
  \citenamefont {Viale},\ and\ \citenamefont {Walczak}}]{Cavagna2015}%
  \BibitemOpen
  \bibfield  {author} {\bibinfo {author} {\bibfnamefont {A.}~\bibnamefont
  {Cavagna}}, \bibinfo {author} {\bibfnamefont {L.}~\bibnamefont
  {Del~Castello}}, \bibinfo {author} {\bibfnamefont {I.}~\bibnamefont
  {Giardina}}, \bibinfo {author} {\bibfnamefont {T.}~\bibnamefont {Grigera}},
  \bibinfo {author} {\bibfnamefont {A.}~\bibnamefont {Jelic}}, \bibinfo
  {author} {\bibfnamefont {S.}~\bibnamefont {Melillo}}, \bibinfo {author}
  {\bibfnamefont {T.}~\bibnamefont {Mora}}, \bibinfo {author} {\bibfnamefont
  {L.}~\bibnamefont {Parisi}}, \bibinfo {author} {\bibfnamefont
  {E.}~\bibnamefont {Silvestri}}, \bibinfo {author} {\bibfnamefont
  {M.}~\bibnamefont {Viale}}, \ and\ \bibinfo {author} {\bibfnamefont {A.~M.}\
  \bibnamefont {Walczak}},\ }\href {\doibase 10.1007/s10955-014-1119-3}
  {\bibfield  {journal} {\bibinfo  {journal} {Journal of Statistical Physics}\
  }\textbf {\bibinfo {volume} {158}},\ \bibinfo {pages} {601} (\bibinfo {year}
  {2015})}\BibitemShut {NoStop}%
\bibitem [{\citenamefont {Young}\ \emph {et~al.}(2013)\citenamefont {Young},
  \citenamefont {Scardovi}, \citenamefont {Cavagna}, \citenamefont {Giardina},\
  and\ \citenamefont {Leonard}}]{Leonard2013}%
  \BibitemOpen
  \bibfield  {author} {\bibinfo {author} {\bibfnamefont {G.~F.}\ \bibnamefont
  {Young}}, \bibinfo {author} {\bibfnamefont {L.}~\bibnamefont {Scardovi}},
  \bibinfo {author} {\bibfnamefont {A.}~\bibnamefont {Cavagna}}, \bibinfo
  {author} {\bibfnamefont {I.}~\bibnamefont {Giardina}}, \ and\ \bibinfo
  {author} {\bibfnamefont {N.~E.}\ \bibnamefont {Leonard}},\ }\href {\doibase
  10.1371/journal.pcbi.1002894} {\bibfield  {journal} {\bibinfo  {journal}
  {PLOS Computational Biology}\ }\textbf {\bibinfo {volume} {9}},\ \bibinfo
  {pages} {1} (\bibinfo {year} {2013})}\BibitemShut {NoStop}%
\bibitem [{\citenamefont {Ballerini}\ \emph {et~al.}(2008)\citenamefont
  {Ballerini}, \citenamefont {Cabibbo}, \citenamefont {Candelier},
  \citenamefont {Cavagna}, \citenamefont {Cisbani}, \citenamefont {Giardina},
  \citenamefont {Lecomte}, \citenamefont {Orlandi}, \citenamefont {Parisi},
  \citenamefont {Procaccini}, \citenamefont {Viale},\ and\ \citenamefont
  {Zdravkovic}}]{Ballerini08}%
  \BibitemOpen
  \bibfield  {author} {\bibinfo {author} {\bibfnamefont {M.}~\bibnamefont
  {Ballerini}}, \bibinfo {author} {\bibfnamefont {N.}~\bibnamefont {Cabibbo}},
  \bibinfo {author} {\bibfnamefont {R.}~\bibnamefont {Candelier}}, \bibinfo
  {author} {\bibfnamefont {A.}~\bibnamefont {Cavagna}}, \bibinfo {author}
  {\bibfnamefont {E.}~\bibnamefont {Cisbani}}, \bibinfo {author} {\bibfnamefont
  {I.}~\bibnamefont {Giardina}}, \bibinfo {author} {\bibfnamefont
  {V.}~\bibnamefont {Lecomte}}, \bibinfo {author} {\bibfnamefont
  {A.}~\bibnamefont {Orlandi}}, \bibinfo {author} {\bibfnamefont
  {G.}~\bibnamefont {Parisi}}, \bibinfo {author} {\bibfnamefont
  {A.}~\bibnamefont {Procaccini}}, \bibinfo {author} {\bibfnamefont
  {M.}~\bibnamefont {Viale}}, \ and\ \bibinfo {author} {\bibfnamefont
  {V.}~\bibnamefont {Zdravkovic}},\ }\href {\doibase 10.1073/pnas.0711437105}
  {\bibfield  {journal} {\bibinfo  {journal} {Proceedings of the National
  Academy of Sciences}\ }\textbf {\bibinfo {volume} {105}},\ \bibinfo {pages}
  {1232} (\bibinfo {year} {2008})},\ \Eprint
  {http://arxiv.org/abs/https://www.pnas.org/content/105/4/1232.full.pdf}
  {https://www.pnas.org/content/105/4/1232.full.pdf} \BibitemShut {NoStop}%
\bibitem [{\citenamefont {Ling}\ \emph {et~al.}(2019)\citenamefont {Ling},
  \citenamefont {E.}, \citenamefont {van~der Vaart~K.}, \citenamefont {T.},
  \citenamefont {A.},\ and\ \citenamefont {Ouellette}}]{Ouellette2019}%
  \BibitemOpen
  \bibfield  {author} {\bibinfo {author} {\bibfnamefont {H.}~\bibnamefont
  {Ling}}, \bibinfo {author} {\bibfnamefont {M.~G.}\ \bibnamefont {E.}},
  \bibinfo {author} {\bibnamefont {van~der Vaart~K.}}, \bibinfo {author}
  {\bibfnamefont {V.~R.}\ \bibnamefont {T.}}, \bibinfo {author} {\bibfnamefont
  {T.}~\bibnamefont {A.}}, \ and\ \bibinfo {author} {\bibfnamefont {N.~T.}\
  \bibnamefont {Ouellette}},\ }\href {\doibase 10.1098/rspb.2019.0865}
  {\bibfield  {journal} {\bibinfo  {journal} {R. Soc. B.}\ }\textbf {\bibinfo
  {volume} {286}} (\bibinfo {year} {2019}),\
  10.1098/rspb.2019.0865}\BibitemShut {NoStop}%
\bibitem [{\citenamefont {Rio}\ and\ \citenamefont
  {Warren}(2014)}]{Rio_Warren_2014}%
  \BibitemOpen
  \bibfield  {author} {\bibinfo {author} {\bibfnamefont {K.}~\bibnamefont
  {Rio}}\ and\ \bibinfo {author} {\bibfnamefont {W.~H.}\ \bibnamefont
  {Warren}},\ }\href {\doibase https://doi.org/10.1016/j.trpro.2014.09.017}
  {\bibfield  {journal} {\bibinfo  {journal} {Transportation Research
  Procedia}\ }\textbf {\bibinfo {volume} {2}},\ \bibinfo {pages} {132 }
  (\bibinfo {year} {2014})},\ \bibinfo {note} {the Conference on Pedestrian and
  Evacuation Dynamics 2014 (PED 2014), 22-24 October 2014, Delft, The
  Netherlands}\BibitemShut {NoStop}%
\bibitem [{\citenamefont {Vicsek}\ and\ \citenamefont
  {Zafeiris}(2012)}]{Vicsek}%
  \BibitemOpen
  \bibfield  {author} {\bibinfo {author} {\bibfnamefont {T.}~\bibnamefont
  {Vicsek}}\ and\ \bibinfo {author} {\bibfnamefont {A.}~\bibnamefont
  {Zafeiris}},\ }\href@noop {} {\bibfield  {journal} {\bibinfo  {journal}
  {Phys. Rep.}\ }\textbf {\bibinfo {volume} {517}},\ \bibinfo {pages} {71}
  (\bibinfo {year} {2012})}\BibitemShut {NoStop}%
\bibitem [{\citenamefont {Marchetti}\ \emph {et~al.}(2013)\citenamefont
  {Marchetti}, \citenamefont {Joanny}, \citenamefont {Ramaswamy}, \citenamefont
  {Liverpool}, \citenamefont {Prost}, \citenamefont {Rao},\ and\ \citenamefont
  {Simha}}]{Marchetti}%
  \BibitemOpen
  \bibfield  {author} {\bibinfo {author} {\bibfnamefont {M.~C.}\ \bibnamefont
  {Marchetti}}, \bibinfo {author} {\bibfnamefont {J.~F.}\ \bibnamefont
  {Joanny}}, \bibinfo {author} {\bibfnamefont {S.}~\bibnamefont {Ramaswamy}},
  \bibinfo {author} {\bibfnamefont {T.~B.}\ \bibnamefont {Liverpool}}, \bibinfo
  {author} {\bibfnamefont {J.}~\bibnamefont {Prost}}, \bibinfo {author}
  {\bibfnamefont {M.}~\bibnamefont {Rao}}, \ and\ \bibinfo {author}
  {\bibfnamefont {R.~A.}\ \bibnamefont {Simha}},\ }\href@noop {} {\bibfield
  {journal} {\bibinfo  {journal} {Rev. Mod. Phys.}\ }\textbf {\bibinfo {volume}
  {85}},\ \bibinfo {pages} {1143} (\bibinfo {year} {2013})}\BibitemShut
  {NoStop}%
\bibitem [{\citenamefont {Aldana}\ \emph {et~al.}(2007)\citenamefont {Aldana},
  \citenamefont {Dossetti}, \citenamefont {Huepe}, \citenamefont {Kenkre},\
  and\ \citenamefont {Larralde}}]{Aldana}%
  \BibitemOpen
  \bibfield  {author} {\bibinfo {author} {\bibfnamefont {M.}~\bibnamefont
  {Aldana}}, \bibinfo {author} {\bibfnamefont {V.}~\bibnamefont {Dossetti}},
  \bibinfo {author} {\bibfnamefont {C.}~\bibnamefont {Huepe}}, \bibinfo
  {author} {\bibfnamefont {V.~M.}\ \bibnamefont {Kenkre}}, \ and\ \bibinfo
  {author} {\bibfnamefont {H.}~\bibnamefont {Larralde}},\ }\href@noop {}
  {\bibfield  {journal} {\bibinfo  {journal} {Phys. Rev. Letts.}\ }\textbf
  {\bibinfo {volume} {98}},\ \bibinfo {pages} {095702} (\bibinfo {year}
  {2007})}\BibitemShut {NoStop}%
\bibitem [{\citenamefont {Chandra}\ \emph {et~al.}(2019)\citenamefont
  {Chandra}, \citenamefont {Girvan},\ and\ \citenamefont
  {Ott}}]{PhysRevX.9.011002}%
  \BibitemOpen
  \bibfield  {author} {\bibinfo {author} {\bibfnamefont {S.}~\bibnamefont
  {Chandra}}, \bibinfo {author} {\bibfnamefont {M.}~\bibnamefont {Girvan}}, \
  and\ \bibinfo {author} {\bibfnamefont {E.}~\bibnamefont {Ott}},\ }\href
  {\doibase 10.1103/PhysRevX.9.011002} {\bibfield  {journal} {\bibinfo
  {journal} {Phys. Rev. X}\ }\textbf {\bibinfo {volume} {9}},\ \bibinfo {pages}
  {011002} (\bibinfo {year} {2019})}\BibitemShut {NoStop}%
\bibitem [{\citenamefont {Solon}\ \emph {et~al.}(2015)\citenamefont {Solon},
  \citenamefont {Fily}, \citenamefont {Baskaran}, \citenamefont {Cates},
  \citenamefont {Kafri}, \citenamefont {Kardar},\ and\ \citenamefont
  {Tailleur}}]{Solon2015}%
  \BibitemOpen
  \bibfield  {author} {\bibinfo {author} {\bibfnamefont {A.}~\bibnamefont
  {Solon}}, \bibinfo {author} {\bibfnamefont {Y.}~\bibnamefont {Fily}},
  \bibinfo {author} {\bibfnamefont {A.}~\bibnamefont {Baskaran}}, \bibinfo
  {author} {\bibfnamefont {M.~E.}\ \bibnamefont {Cates}}, \bibinfo {author}
  {\bibfnamefont {Y.}~\bibnamefont {Kafri}}, \bibinfo {author} {\bibfnamefont
  {M.}~\bibnamefont {Kardar}}, \ and\ \bibinfo {author} {\bibfnamefont
  {J.}~\bibnamefont {Tailleur}},\ }\href {\doibase 10.1038/nphys3377}
  {\bibfield  {journal} {\bibinfo  {journal} {Nature Phys.}\ }\textbf {\bibinfo
  {volume} {11}},\ \bibinfo {pages} {673} (\bibinfo {year} {2015})}\BibitemShut
  {NoStop}%
\bibitem [{\citenamefont {Fodor}\ \emph {et~al.}(2016)\citenamefont {Fodor},
  \citenamefont {Nardini}, \citenamefont {Cates}, \citenamefont {Tailleur},
  \citenamefont {Visco},\ and\ \citenamefont {van
  Wijland}}]{PhysRevLett.117.038103}%
  \BibitemOpen
  \bibfield  {author} {\bibinfo {author} {\bibfnamefont {E.}~\bibnamefont
  {Fodor}}, \bibinfo {author} {\bibfnamefont {C.}~\bibnamefont {Nardini}},
  \bibinfo {author} {\bibfnamefont {M.~E.}\ \bibnamefont {Cates}}, \bibinfo
  {author} {\bibfnamefont {J.}~\bibnamefont {Tailleur}}, \bibinfo {author}
  {\bibfnamefont {P.}~\bibnamefont {Visco}}, \ and\ \bibinfo {author}
  {\bibfnamefont {F.}~\bibnamefont {van Wijland}},\ }\href {\doibase
  10.1103/PhysRevLett.117.038103} {\bibfield  {journal} {\bibinfo  {journal}
  {Phys. Rev. Lett.}\ }\textbf {\bibinfo {volume} {117}},\ \bibinfo {pages}
  {038103} (\bibinfo {year} {2016})}\BibitemShut {NoStop}%
\bibitem [{\citenamefont {Woodhouse}\ \emph {et~al.}(2018)\citenamefont
  {Woodhouse}, \citenamefont {Ronellenfitsch},\ and\ \citenamefont
  {Dunkel}}]{PhysRevLett.121.178001}%
  \BibitemOpen
  \bibfield  {author} {\bibinfo {author} {\bibfnamefont {F.~G.}\ \bibnamefont
  {Woodhouse}}, \bibinfo {author} {\bibfnamefont {H.}~\bibnamefont
  {Ronellenfitsch}}, \ and\ \bibinfo {author} {\bibfnamefont {J.}~\bibnamefont
  {Dunkel}},\ }\href {\doibase 10.1103/PhysRevLett.121.178001} {\bibfield
  {journal} {\bibinfo  {journal} {Phys. Rev. Lett.}\ }\textbf {\bibinfo
  {volume} {121}},\ \bibinfo {pages} {178001} (\bibinfo {year}
  {2018})}\BibitemShut {NoStop}%
\bibitem [{\citenamefont {Woillez}\ \emph {et~al.}(2019)\citenamefont
  {Woillez}, \citenamefont {Zhao}, \citenamefont {Kafri}, \citenamefont
  {Lecomte},\ and\ \citenamefont {Tailleur}}]{PhysRevLett.122.258001}%
  \BibitemOpen
  \bibfield  {author} {\bibinfo {author} {\bibfnamefont {E.}~\bibnamefont
  {Woillez}}, \bibinfo {author} {\bibfnamefont {Y.}~\bibnamefont {Zhao}},
  \bibinfo {author} {\bibfnamefont {Y.}~\bibnamefont {Kafri}}, \bibinfo
  {author} {\bibfnamefont {V.}~\bibnamefont {Lecomte}}, \ and\ \bibinfo
  {author} {\bibfnamefont {J.}~\bibnamefont {Tailleur}},\ }\href {\doibase
  10.1103/PhysRevLett.122.258001} {\bibfield  {journal} {\bibinfo  {journal}
  {Phys. Rev. Lett.}\ }\textbf {\bibinfo {volume} {122}},\ \bibinfo {pages}
  {258001} (\bibinfo {year} {2019})}\BibitemShut {NoStop}%
\bibitem [{\citenamefont {{Desai}}\ \emph {et~al.}(2001)\citenamefont
  {{Desai}}, \citenamefont {{Ostrowski}},\ and\ \citenamefont
  {{Kumar}}}]{Desai01}%
  \BibitemOpen
  \bibfield  {author} {\bibinfo {author} {\bibfnamefont {J.~P.}\ \bibnamefont
  {{Desai}}}, \bibinfo {author} {\bibfnamefont {J.~P.}\ \bibnamefont
  {{Ostrowski}}}, \ and\ \bibinfo {author} {\bibfnamefont {V.}~\bibnamefont
  {{Kumar}}},\ }in\ \href@noop {} {\emph {\bibinfo {booktitle} {IEEE
  Transactions on Robotics and Automation}}},\ Vol.\ \bibinfo {volume} {17(6)}\
  (\bibinfo {year} {2001})\ pp.\ \bibinfo {pages} {905--908}\BibitemShut
  {NoStop}%
\bibitem [{\citenamefont {{Jadbabaie}}\ \emph {et~al.}(2003)\citenamefont
  {{Jadbabaie}}, \citenamefont {{Jie Lin}},\ and\ \citenamefont
  {{Morse}}}]{Jadbabaie03}%
  \BibitemOpen
  \bibfield  {author} {\bibinfo {author} {\bibfnamefont {A.}~\bibnamefont
  {{Jadbabaie}}}, \bibinfo {author} {\bibnamefont {{Jie Lin}}}, \ and\ \bibinfo
  {author} {\bibfnamefont {A.~S.}\ \bibnamefont {{Morse}}},\ }\href {\doibase
  10.1109/TAC.2003.812781} {\bibfield  {journal} {\bibinfo  {journal} {IEEE
  Transactions on Automatic Control}\ }\textbf {\bibinfo {volume} {48}},\
  \bibinfo {pages} {988} (\bibinfo {year} {2003})}\BibitemShut {NoStop}%
\bibitem [{\citenamefont {{Tanner}}\ \emph
  {et~al.}(2003{\natexlab{a}})\citenamefont {{Tanner}}, \citenamefont
  {{Jadbabaie}},\ and\ \citenamefont {{Pappas}}}]{Tanner03b}%
  \BibitemOpen
  \bibfield  {author} {\bibinfo {author} {\bibfnamefont {H.~G.}\ \bibnamefont
  {{Tanner}}}, \bibinfo {author} {\bibfnamefont {A.}~\bibnamefont
  {{Jadbabaie}}}, \ and\ \bibinfo {author} {\bibfnamefont {G.~J.}\ \bibnamefont
  {{Pappas}}},\ }in\ \href {\doibase 10.1109/CDC.2003.1272911} {\emph {\bibinfo
  {booktitle} {42nd IEEE International Conference on Decision and Control (IEEE
  Cat. No.03CH37475)}}},\ Vol.~\bibinfo {volume} {2}\ (\bibinfo {year} {2003})\
  pp.\ \bibinfo {pages} {2016--2021 Vol.2}\BibitemShut {NoStop}%
\bibitem [{\citenamefont {{Tanner}}\ \emph
  {et~al.}(2003{\natexlab{b}})\citenamefont {{Tanner}}, \citenamefont
  {{Jadbabaie}},\ and\ \citenamefont {{Pappas}}}]{Tanner03a}%
  \BibitemOpen
  \bibfield  {author} {\bibinfo {author} {\bibfnamefont {H.~G.}\ \bibnamefont
  {{Tanner}}}, \bibinfo {author} {\bibfnamefont {A.}~\bibnamefont
  {{Jadbabaie}}}, \ and\ \bibinfo {author} {\bibfnamefont {G.~J.}\ \bibnamefont
  {{Pappas}}},\ }in\ \href {\doibase 10.1109/CDC.2003.1272910} {\emph {\bibinfo
  {booktitle} {42nd IEEE International Conference on Decision and Control (IEEE
  Cat. No.03CH37475)}}},\ Vol.~\bibinfo {volume} {2}\ (\bibinfo {year} {2003})\
  pp.\ \bibinfo {pages} {2010--2015 Vol.2}\BibitemShut {NoStop}%
\bibitem [{\citenamefont {{Gazi}}(2005)}]{Gazi05}%
  \BibitemOpen
  \bibfield  {author} {\bibinfo {author} {\bibfnamefont {V.}~\bibnamefont
  {{Gazi}}},\ }\href {\doibase 10.1109/TRO.2005.853487} {\bibfield  {journal}
  {\bibinfo  {journal} {IEEE Transactions on Robotics}\ }\textbf {\bibinfo
  {volume} {21}},\ \bibinfo {pages} {1208} (\bibinfo {year}
  {2005})}\BibitemShut {NoStop}%
\bibitem [{\citenamefont {{Tanner}}\ \emph {et~al.}(2007)\citenamefont
  {{Tanner}}, \citenamefont {{Jadbabaie}},\ and\ \citenamefont
  {{Pappas}}}]{Tanner07}%
  \BibitemOpen
  \bibfield  {author} {\bibinfo {author} {\bibfnamefont {H.~G.}\ \bibnamefont
  {{Tanner}}}, \bibinfo {author} {\bibfnamefont {A.}~\bibnamefont
  {{Jadbabaie}}}, \ and\ \bibinfo {author} {\bibfnamefont {G.~J.}\ \bibnamefont
  {{Pappas}}},\ }\href {\doibase 10.1109/TAC.2007.895948} {\bibfield  {journal}
  {\bibinfo  {journal} {IEEE Transactions on Automatic Control}\ }\textbf
  {\bibinfo {volume} {52}},\ \bibinfo {pages} {863} (\bibinfo {year}
  {2007})}\BibitemShut {NoStop}%
\bibitem [{\citenamefont {Ramachandran}\ \emph {et~al.}(2018)\citenamefont
  {Ramachandran}, \citenamefont {Elamvazhuthi},\ and\ \citenamefont
  {Berman}}]{Ramachandran2018}%
  \BibitemOpen
  \bibfield  {author} {\bibinfo {author} {\bibfnamefont {R.~K.}\ \bibnamefont
  {Ramachandran}}, \bibinfo {author} {\bibfnamefont {K.}~\bibnamefont
  {Elamvazhuthi}}, \ and\ \bibinfo {author} {\bibfnamefont {S.}~\bibnamefont
  {Berman}},\ }\enquote {\bibinfo {title} {An optimal control approach to
  mapping gps-denied environments using a stochastic robotic swarm},}\ in\
  \href {\doibase 10.1007/978-3-319-51532-8_29} {\emph {\bibinfo {booktitle}
  {Robotics Research: Volume 1}}},\ \bibinfo {editor} {edited by\ \bibinfo
  {editor} {\bibfnamefont {A.}~\bibnamefont {Bicchi}}\ and\ \bibinfo {editor}
  {\bibfnamefont {W.}~\bibnamefont {Burgard}}}\ (\bibinfo  {publisher}
  {Springer International Publishing},\ \bibinfo {address} {Cham},\ \bibinfo
  {year} {2018})\ pp.\ \bibinfo {pages} {477--493}\BibitemShut {NoStop}%
\bibitem [{\citenamefont {{Li}}\ \emph {et~al.}(2017)\citenamefont {{Li}},
  \citenamefont {{Feng}}, \citenamefont {{Ehrhard}}, \citenamefont {{Shen}},
  \citenamefont {{Cobos}}, \citenamefont {{Zhang}}, \citenamefont
  {{Elamvazhuthi}}, \citenamefont {{Berman}}, \citenamefont {{Haberland}},\
  and\ \citenamefont {{Bertozzi}}}]{Li17}%
  \BibitemOpen
  \bibfield  {author} {\bibinfo {author} {\bibfnamefont {H.}~\bibnamefont
  {{Li}}}, \bibinfo {author} {\bibfnamefont {C.}~\bibnamefont {{Feng}}},
  \bibinfo {author} {\bibfnamefont {H.}~\bibnamefont {{Ehrhard}}}, \bibinfo
  {author} {\bibfnamefont {Y.}~\bibnamefont {{Shen}}}, \bibinfo {author}
  {\bibfnamefont {B.}~\bibnamefont {{Cobos}}}, \bibinfo {author} {\bibfnamefont
  {F.}~\bibnamefont {{Zhang}}}, \bibinfo {author} {\bibfnamefont
  {K.}~\bibnamefont {{Elamvazhuthi}}}, \bibinfo {author} {\bibfnamefont
  {S.}~\bibnamefont {{Berman}}}, \bibinfo {author} {\bibfnamefont
  {M.}~\bibnamefont {{Haberland}}}, \ and\ \bibinfo {author} {\bibfnamefont
  {A.~L.}\ \bibnamefont {{Bertozzi}}},\ }in\ \href {\doibase
  10.1109/IROS.2017.8206299} {\emph {\bibinfo {booktitle} {2017 IEEE/RSJ
  International Conference on Intelligent Robots and Systems (IROS)}}}\
  (\bibinfo {year} {2017})\ pp.\ \bibinfo {pages} {4341--4347}\BibitemShut
  {NoStop}%
\bibitem [{\citenamefont {{Berman}}\ \emph {et~al.}(2007)\citenamefont
  {{Berman}}, \citenamefont {{Halasz}}, \citenamefont {{Kumar}},\ and\
  \citenamefont {{Pratt}}}]{Berman07}%
  \BibitemOpen
  \bibfield  {author} {\bibinfo {author} {\bibfnamefont {S.}~\bibnamefont
  {{Berman}}}, \bibinfo {author} {\bibfnamefont {A.}~\bibnamefont {{Halasz}}},
  \bibinfo {author} {\bibfnamefont {V.}~\bibnamefont {{Kumar}}}, \ and\
  \bibinfo {author} {\bibfnamefont {S.}~\bibnamefont {{Pratt}}},\ }in\ \href
  {\doibase 10.1109/ROBOT.2007.363665} {\emph {\bibinfo {booktitle}
  {Proceedings 2007 IEEE International Conference on Robotics and
  Automation}}}\ (\bibinfo {year} {2007})\ pp.\ \bibinfo {pages}
  {2318--2323}\BibitemShut {NoStop}%
\bibitem [{\citenamefont {Hsieh}\ \emph {et~al.}(2008)\citenamefont {Hsieh},
  \citenamefont {Hal{\'a}sz}, \citenamefont {Berman},\ and\ \citenamefont
  {Kumar}}]{Hsieh2008}%
  \BibitemOpen
  \bibfield  {author} {\bibinfo {author} {\bibfnamefont {M.~A.}\ \bibnamefont
  {Hsieh}}, \bibinfo {author} {\bibfnamefont {{\'A}.}~\bibnamefont
  {Hal{\'a}sz}}, \bibinfo {author} {\bibfnamefont {S.}~\bibnamefont {Berman}},
  \ and\ \bibinfo {author} {\bibfnamefont {V.}~\bibnamefont {Kumar}},\ }\href
  {\doibase 10.1007/s11721-008-0019-z} {\bibfield  {journal} {\bibinfo
  {journal} {Swarm Intelligence}\ }\textbf {\bibinfo {volume} {2}},\ \bibinfo
  {pages} {121} (\bibinfo {year} {2008})}\BibitemShut {NoStop}%
\bibitem [{\citenamefont {Wiech}\ \emph {et~al.}(2018)\citenamefont {Wiech},
  \citenamefont {Eremeyev},\ and\ \citenamefont {Giorgio}}]{Wiech2018}%
  \BibitemOpen
  \bibfield  {author} {\bibinfo {author} {\bibfnamefont {J.}~\bibnamefont
  {Wiech}}, \bibinfo {author} {\bibfnamefont {V.~A.}\ \bibnamefont {Eremeyev}},
  \ and\ \bibinfo {author} {\bibfnamefont {I.}~\bibnamefont {Giorgio}},\ }\href
  {\doibase 10.1007/s00161-018-0664-4} {\bibfield  {journal} {\bibinfo
  {journal} {Continuum Mechanics and Thermodynamics}\ }\textbf {\bibinfo
  {volume} {30}},\ \bibinfo {pages} {1091} (\bibinfo {year}
  {2018})}\BibitemShut {NoStop}%
\bibitem [{\citenamefont {Keller}\ \emph {et~al.}(2008)\citenamefont {Keller},
  \citenamefont {Schmidt}, \citenamefont {Wittbrodt},\ and\ \citenamefont
  {Stelzer}}]{Science2008}%
  \BibitemOpen
  \bibfield  {author} {\bibinfo {author} {\bibfnamefont {P.~J.}\ \bibnamefont
  {Keller}}, \bibinfo {author} {\bibfnamefont {A.~D.}\ \bibnamefont {Schmidt}},
  \bibinfo {author} {\bibfnamefont {J.}~\bibnamefont {Wittbrodt}}, \ and\
  \bibinfo {author} {\bibfnamefont {E.~H.~K.}\ \bibnamefont {Stelzer}},\
  }\href@noop {} {\bibfield  {journal} {\bibinfo  {journal} {Science}\ }\textbf
  {\bibinfo {volume} {322}},\ \bibinfo {pages} {1065} (\bibinfo {year}
  {2008})}\BibitemShut {NoStop}%
\bibitem [{\citenamefont {Collinson}\ \emph {et~al.}(2002)\citenamefont
  {Collinson}, \citenamefont {Morris}, \citenamefont {Reid}, \citenamefont
  {Ramaesh}, \citenamefont {Keighren}, \citenamefont {Flockhart}, \citenamefont
  {Hill}, \citenamefont {Tan}, \citenamefont {Ramaesh}, \citenamefont
  {Dhillon},\ and\ \citenamefont {West}}]{CornealGrowth}%
  \BibitemOpen
  \bibfield  {author} {\bibinfo {author} {\bibfnamefont {J.}~\bibnamefont
  {Collinson}}, \bibinfo {author} {\bibfnamefont {L.}~\bibnamefont {Morris}},
  \bibinfo {author} {\bibfnamefont {A.}~\bibnamefont {Reid}}, \bibinfo {author}
  {\bibfnamefont {T.}~\bibnamefont {Ramaesh}}, \bibinfo {author} {\bibfnamefont
  {M.}~\bibnamefont {Keighren}}, \bibinfo {author} {\bibfnamefont
  {J.}~\bibnamefont {Flockhart}}, \bibinfo {author} {\bibfnamefont
  {R.}~\bibnamefont {Hill}}, \bibinfo {author} {\bibfnamefont {S.}~\bibnamefont
  {Tan}}, \bibinfo {author} {\bibfnamefont {K.}~\bibnamefont {Ramaesh}},
  \bibinfo {author} {\bibfnamefont {B.}~\bibnamefont {Dhillon}}, \ and\
  \bibinfo {author} {\bibfnamefont {J.}~\bibnamefont {West}},\ }\href@noop {}
  {\bibfield  {journal} {\bibinfo  {journal} {Dev. Dynam.}\ }\textbf {\bibinfo
  {volume} {224}},\ \bibinfo {pages} {432} (\bibinfo {year}
  {2002})}\BibitemShut {NoStop}%
\bibitem [{\citenamefont {Keber}\ \emph {et~al.}(2014)\citenamefont {Keber},
  \citenamefont {Loiseau}, \citenamefont {Sanchez}, \citenamefont {DeCamp},
  \citenamefont {Giomi}, \citenamefont {Bowick}, \citenamefont {Marchetti},
  \citenamefont {Dogic},\ and\ \citenamefont {Bausch}}]{Topology2014}%
  \BibitemOpen
  \bibfield  {author} {\bibinfo {author} {\bibfnamefont {F.}~\bibnamefont
  {Keber}}, \bibinfo {author} {\bibfnamefont {E.}~\bibnamefont {Loiseau}},
  \bibinfo {author} {\bibfnamefont {T.}~\bibnamefont {Sanchez}}, \bibinfo
  {author} {\bibfnamefont {S.}~\bibnamefont {DeCamp}}, \bibinfo {author}
  {\bibfnamefont {L.}~\bibnamefont {Giomi}}, \bibinfo {author} {\bibfnamefont
  {M.}~\bibnamefont {Bowick}}, \bibinfo {author} {\bibfnamefont
  {M.}~\bibnamefont {Marchetti}}, \bibinfo {author} {\bibfnamefont
  {Z.}~\bibnamefont {Dogic}}, \ and\ \bibinfo {author} {\bibfnamefont
  {A.}~\bibnamefont {Bausch}},\ }\href@noop {} {\bibfield  {journal} {\bibinfo
  {journal} {Science}\ }\textbf {\bibinfo {volume} {345}},\ \bibinfo {pages}
  {1135} (\bibinfo {year} {2014})}\BibitemShut {NoStop}%
\bibitem [{\citenamefont {Zhang}\ \emph {et~al.}(2016)\citenamefont {Zhang},
  \citenamefont {Zhou}, \citenamefont {Rahimi},\ and\ \citenamefont
  {de~Pablo}}]{NematicShells}%
  \BibitemOpen
  \bibfield  {author} {\bibinfo {author} {\bibfnamefont {R.}~\bibnamefont
  {Zhang}}, \bibinfo {author} {\bibfnamefont {Y.}~\bibnamefont {Zhou}},
  \bibinfo {author} {\bibfnamefont {M.}~\bibnamefont {Rahimi}}, \ and\ \bibinfo
  {author} {\bibfnamefont {J.~J.}\ \bibnamefont {de~Pablo}},\ }\href@noop {}
  {\bibfield  {journal} {\bibinfo  {journal} {Nat. Commun.}\ }\textbf {\bibinfo
  {volume} {7}},\ \bibinfo {pages} {13483} (\bibinfo {year}
  {2016})}\BibitemShut {NoStop}%
\bibitem [{\citenamefont {Markdahl}\ \emph {et~al.}(2018)\citenamefont
  {Markdahl}, \citenamefont {Thunberg},\ and\ \citenamefont
  {Gonalves}}]{ConsensusOnSphere}%
  \BibitemOpen
  \bibfield  {author} {\bibinfo {author} {\bibfnamefont {J.}~\bibnamefont
  {Markdahl}}, \bibinfo {author} {\bibfnamefont {J.}~\bibnamefont {Thunberg}},
  \ and\ \bibinfo {author} {\bibfnamefont {J.}~\bibnamefont {Gonalves}},\
  }\href@noop {} {\bibfield  {journal} {\bibinfo  {journal} {IEEE Transactions
  on Automatic Control}\ }\textbf {\bibinfo {volume} {63}},\ \bibinfo {pages}
  {1664} (\bibinfo {year} {2018})}\BibitemShut {NoStop}%
\bibitem [{\citenamefont {Sknepnek}\ and\ \citenamefont
  {Henkes}(2015)}]{PhysRevE.91.022306}%
  \BibitemOpen
  \bibfield  {author} {\bibinfo {author} {\bibfnamefont {R.}~\bibnamefont
  {Sknepnek}}\ and\ \bibinfo {author} {\bibfnamefont {S.}~\bibnamefont
  {Henkes}},\ }\href {\doibase 10.1103/PhysRevE.91.022306} {\bibfield
  {journal} {\bibinfo  {journal} {Phys. Rev. E}\ }\textbf {\bibinfo {volume}
  {91}},\ \bibinfo {pages} {022306} (\bibinfo {year} {2015})}\BibitemShut
  {NoStop}%
\bibitem [{\citenamefont {Li}(2015)}]{Li2015}%
  \BibitemOpen
  \bibfield  {author} {\bibinfo {author} {\bibfnamefont {W.}~\bibnamefont
  {Li}},\ }\href {\doibase 10.1038/srep13603} {\bibfield  {journal} {\bibinfo
  {journal} {Sci. Rep.}\ }\textbf {\bibinfo {volume} {5}},\ \bibinfo {pages}
  {13603} (\bibinfo {year} {2015})}\BibitemShut {NoStop}%
\bibitem [{\citenamefont {Praetorius}\ \emph {et~al.}(2018)\citenamefont
  {Praetorius}, \citenamefont {Voigt}, \citenamefont {Wittkowski},\ and\
  \citenamefont {L\"owen}}]{PhysRevE.97.052615}%
  \BibitemOpen
  \bibfield  {author} {\bibinfo {author} {\bibfnamefont {S.}~\bibnamefont
  {Praetorius}}, \bibinfo {author} {\bibfnamefont {A.}~\bibnamefont {Voigt}},
  \bibinfo {author} {\bibfnamefont {R.}~\bibnamefont {Wittkowski}}, \ and\
  \bibinfo {author} {\bibfnamefont {H.}~\bibnamefont {L\"owen}},\ }\href
  {\doibase 10.1103/PhysRevE.97.052615} {\bibfield  {journal} {\bibinfo
  {journal} {Phys. Rev. E}\ }\textbf {\bibinfo {volume} {97}},\ \bibinfo
  {pages} {052615} (\bibinfo {year} {2018})}\BibitemShut {NoStop}%
\bibitem [{\citenamefont {Janssen}\ \emph {et~al.}(2015)\citenamefont
  {Janssen}, \citenamefont {Kaise},\ and\ \citenamefont
  {L{\"o}wen}}]{Janssen2017}%
  \BibitemOpen
  \bibfield  {author} {\bibinfo {author} {\bibfnamefont {L.~M.~C.}\
  \bibnamefont {Janssen}}, \bibinfo {author} {\bibfnamefont {A.}~\bibnamefont
  {Kaise}}, \ and\ \bibinfo {author} {\bibfnamefont {H.}~\bibnamefont
  {L{\"o}wen}},\ }\href {\doibase 10.1038/srep13603} {\bibfield  {journal}
  {\bibinfo  {journal} {Sci. Rep.}\ }\textbf {\bibinfo {volume} {5}},\ \bibinfo
  {pages} {13603} (\bibinfo {year} {2015})}\BibitemShut {NoStop}%
\bibitem [{\citenamefont {Apazaa}\ and\ \citenamefont
  {Sandoval}(2018)}]{Riemannian2018}%
  \BibitemOpen
  \bibfield  {author} {\bibinfo {author} {\bibfnamefont {L.}~\bibnamefont
  {Apazaa}}\ and\ \bibinfo {author} {\bibfnamefont {M.}~\bibnamefont
  {Sandoval}},\ }\href {\doibase 10.1039/C8SM01034J} {\bibfield  {journal}
  {\bibinfo  {journal} {Soft Matter}\ }\textbf {\bibinfo {volume} {14}},\
  \bibinfo {pages} {9928} (\bibinfo {year} {2018})}\BibitemShut {NoStop}%
\bibitem [{\citenamefont {Castro-Villarreal}\ and\ \citenamefont
  {Sevilla}(2018)}]{PhysRevE.97.052605}%
  \BibitemOpen
  \bibfield  {author} {\bibinfo {author} {\bibfnamefont {P.}~\bibnamefont
  {Castro-Villarreal}}\ and\ \bibinfo {author} {\bibfnamefont {F.~J.}\
  \bibnamefont {Sevilla}},\ }\href {\doibase 10.1103/PhysRevE.97.052605}
  {\bibfield  {journal} {\bibinfo  {journal} {Phys. Rev. E}\ }\textbf {\bibinfo
  {volume} {97}},\ \bibinfo {pages} {052605} (\bibinfo {year}
  {2018})}\BibitemShut {NoStop}%
\bibitem [{\citenamefont {Gazi}\ and\ \citenamefont
  {Passino}(2003)}]{GaziStabilityBook}%
  \BibitemOpen
  \bibfield  {author} {\bibinfo {author} {\bibfnamefont {V.}~\bibnamefont
  {Gazi}}\ and\ \bibinfo {author} {\bibfnamefont {K.~M.}\ \bibnamefont
  {Passino}},\ }\href {\doibase 10.1007/978-3-642-18041-5} {\emph {\bibinfo
  {title} {Swarm Stability and Optimization}}}\ (\bibinfo  {publisher}
  {Springer, New York},\ \bibinfo {year} {2003})\BibitemShut {NoStop}%
\bibitem [{\citenamefont {Albi}\ \emph {et~al.}(2014)\citenamefont {Albi},
  \citenamefont {BalaguŽ}, \citenamefont {Carrillo},\ and\ \citenamefont {von
  Brecht}}]{AlbiStability2014}%
  \BibitemOpen
  \bibfield  {author} {\bibinfo {author} {\bibfnamefont {G.}~\bibnamefont
  {Albi}}, \bibinfo {author} {\bibfnamefont {D.}~\bibnamefont {BalaguŽ}},
  \bibinfo {author} {\bibfnamefont {J.~A.}\ \bibnamefont {Carrillo}}, \ and\
  \bibinfo {author} {\bibfnamefont {J.}~\bibnamefont {von Brecht}},\ }\href
  {\doibase 10.1137/13091779X} {\bibfield  {journal} {\bibinfo  {journal} {SIAM
  J. Appl. Math.}\ }\textbf {\bibinfo {volume} {74}},\ \bibinfo {pages} {794}
  (\bibinfo {year} {2014})}\BibitemShut {NoStop}%
\bibitem [{\citenamefont {Hindes}\ \emph {et~al.}(2020)\citenamefont {Hindes},
  \citenamefont {Edwards}, \citenamefont {Kamimoto}, \citenamefont {Triandaf},\
  and\ \citenamefont {Schwartz}}]{PhysRevE.101.042202}%
  \BibitemOpen
  \bibfield  {author} {\bibinfo {author} {\bibfnamefont {J.}~\bibnamefont
  {Hindes}}, \bibinfo {author} {\bibfnamefont {V.}~\bibnamefont {Edwards}},
  \bibinfo {author} {\bibfnamefont {S.}~\bibnamefont {Kamimoto}}, \bibinfo
  {author} {\bibfnamefont {I.}~\bibnamefont {Triandaf}}, \ and\ \bibinfo
  {author} {\bibfnamefont {I.~B.}\ \bibnamefont {Schwartz}},\ }\href {\doibase
  10.1103/PhysRevE.101.042202} {\bibfield  {journal} {\bibinfo  {journal}
  {Phys. Rev. E}\ }\textbf {\bibinfo {volume} {101}},\ \bibinfo {pages}
  {042202} (\bibinfo {year} {2020})}\BibitemShut {NoStop}%
\bibitem [{\citenamefont {Levine}\ \emph {et~al.}(2000)\citenamefont {Levine},
  \citenamefont {Rappel},\ and\ \citenamefont {Cohen}}]{Levine}%
  \BibitemOpen
  \bibfield  {author} {\bibinfo {author} {\bibfnamefont {H.}~\bibnamefont
  {Levine}}, \bibinfo {author} {\bibfnamefont {W.~J.}\ \bibnamefont {Rappel}},
  \ and\ \bibinfo {author} {\bibfnamefont {I.}~\bibnamefont {Cohen}},\
  }\href@noop {} {\bibfield  {journal} {\bibinfo  {journal} {Phys. Rev. E}\
  }\textbf {\bibinfo {volume} {63}},\ \bibinfo {pages} {017101} (\bibinfo
  {year} {2000})}\BibitemShut {NoStop}%
\bibitem [{\citenamefont {Erdmann}\ \emph {et~al.}(2005)\citenamefont
  {Erdmann}, \citenamefont {Ebeling},\ and\ \citenamefont
  {Mikhailov}}]{Erdmann}%
  \BibitemOpen
  \bibfield  {author} {\bibinfo {author} {\bibfnamefont {U.}~\bibnamefont
  {Erdmann}}, \bibinfo {author} {\bibfnamefont {W.}~\bibnamefont {Ebeling}}, \
  and\ \bibinfo {author} {\bibfnamefont {A.~S.}\ \bibnamefont {Mikhailov}},\
  }\href@noop {} {\bibfield  {journal} {\bibinfo  {journal} {Phys. Rev. E}\
  }\textbf {\bibinfo {volume} {71}},\ \bibinfo {pages} {051904} (\bibinfo
  {year} {2005})}\BibitemShut {NoStop}%
\bibitem [{\citenamefont {Minguzzi}(2015)}]{Minguzzi}%
  \BibitemOpen
  \bibfield  {author} {\bibinfo {author} {\bibfnamefont {E.}~\bibnamefont
  {Minguzzi}},\ }\href@noop {} {\bibfield  {journal} {\bibinfo  {journal}
  {European Journal of Physics}\ }\textbf {\bibinfo {volume} {36}},\ \bibinfo
  {pages} {035014} (\bibinfo {year} {2015})}\BibitemShut {NoStop}%
\bibitem [{\citenamefont {D'Orsogna}\ \emph {et~al.}(2006)\citenamefont
  {D'Orsogna}, \citenamefont {Chuang}, \citenamefont {Bertozzi},\ and\
  \citenamefont {Chayes}}]{DOrsagna}%
  \BibitemOpen
  \bibfield  {author} {\bibinfo {author} {\bibfnamefont {M.~R.}\ \bibnamefont
  {D'Orsogna}}, \bibinfo {author} {\bibfnamefont {Y.~L.}\ \bibnamefont
  {Chuang}}, \bibinfo {author} {\bibfnamefont {A.~L.}\ \bibnamefont
  {Bertozzi}}, \ and\ \bibinfo {author} {\bibfnamefont {L.~S.}\ \bibnamefont
  {Chayes}},\ }\href@noop {} {\bibfield  {journal} {\bibinfo  {journal} {Phys.
  Rev. Lett.}\ }\textbf {\bibinfo {volume} {96}},\ \bibinfo {pages} {104302}
  (\bibinfo {year} {2006})}\BibitemShut {NoStop}%
\bibitem [{\citenamefont {Edwards}\ \emph {et~al.}()\citenamefont {Edwards},
  \citenamefont {deZonia}, \citenamefont {Hsieh}, \citenamefont {Hindes},
  \citenamefont {Triandof},\ and\ \citenamefont {Schwartz}}]{Edwards2019}%
  \BibitemOpen
  \bibinfo {author} {\bibfnamefont {V.}~\bibnamefont {Edwards}}, \bibinfo
  {author} {\bibfnamefont {P.}~\bibnamefont {deZonia}}, \bibinfo {author}
  {\bibfnamefont {M.~A.}\ \bibnamefont {Hsieh}}, \bibinfo {author}
  {\bibfnamefont {J.}~\bibnamefont {Hindes}}, \bibinfo {author} {\bibfnamefont
  {I.}~\bibnamefont {Triandof}}, \ and\ \bibinfo {author} {\bibfnamefont
  {I.~B.}\ \bibnamefont {Schwartz}}\BibitemShut {NoStop}%
\bibitem [{\citenamefont {Szwaykowska}\ \emph {et~al.}(2016)\citenamefont
  {Szwaykowska}, \citenamefont {Schwartz}, \citenamefont {Mier-y Teran~Romero},
  \citenamefont {Heckman}, \citenamefont {Mox},\ and\ \citenamefont
  {Hsieh}}]{Szwaykowska2016}%
  \BibitemOpen
\bibfield  {author} {  }\bibfield  {author} {\bibinfo {author} {\bibfnamefont
  {K.}~\bibnamefont {Szwaykowska}}, \bibinfo {author} {\bibfnamefont {I.~B.}\
  \bibnamefont {Schwartz}}, \bibinfo {author} {\bibfnamefont {L.}~\bibnamefont
  {Mier-y Teran~Romero}}, \bibinfo {author} {\bibfnamefont {C.~R.}\
  \bibnamefont {Heckman}}, \bibinfo {author} {\bibfnamefont {D.}~\bibnamefont
  {Mox}}, \ and\ \bibinfo {author} {\bibfnamefont {M.~A.}\ \bibnamefont
  {Hsieh}},\ }\href {\doibase 10.1103/PhysRevE.93.032307} {\bibfield  {journal}
  {\bibinfo  {journal} {Phys. Rev. E}\ }\textbf {\bibinfo {volume} {93}},\
  \bibinfo {pages} {032307} (\bibinfo {year} {2016})}\BibitemShut {NoStop}%
\bibitem [{\citenamefont {Forgoston}\ and\ \citenamefont
  {Schwartz}(2008)}]{F1}%
  \BibitemOpen
  \bibfield  {author} {\bibinfo {author} {\bibfnamefont {E.}~\bibnamefont
  {Forgoston}}\ and\ \bibinfo {author} {\bibfnamefont {I.~B.}\ \bibnamefont
  {Schwartz}},\ }\href@noop {} {\bibfield  {journal} {\bibinfo  {journal}
  {Phys. Rev. E}\ }\textbf {\bibinfo {volume} {77}},\ \bibinfo {pages}
  {035203(R)} (\bibinfo {year} {2008})}\BibitemShut {NoStop}%
\bibitem [{\citenamefont {y~Teran-Romero}\ \emph {et~al.}(2012)\citenamefont
  {y~Teran-Romero}, \citenamefont {Forgoston},\ and\ \citenamefont
  {Schwartz}}]{Romero2012}%
  \BibitemOpen
  \bibfield  {author} {\bibinfo {author} {\bibfnamefont {L.~M.}\ \bibnamefont
  {y~Teran-Romero}}, \bibinfo {author} {\bibfnamefont {E.}~\bibnamefont
  {Forgoston}}, \ and\ \bibinfo {author} {\bibfnamefont {I.~B.}\ \bibnamefont
  {Schwartz}},\ }\href {\doibase 10.1109/TRO.2012.2198511} {\bibfield
  {journal} {\bibinfo  {journal} {IEEE Transactions on Robotics}\ }\textbf
  {\bibinfo {volume} {28}},\ \bibinfo {pages} {1034} (\bibinfo {year}
  {2012})}\BibitemShut {NoStop}%
\bibitem [{\citenamefont {Hindes}\ \emph {et~al.}(2016)\citenamefont {Hindes},
  \citenamefont {Szwaykowska},\ and\ \citenamefont {Schwartz}}]{J1}%
  \BibitemOpen
  \bibfield  {author} {\bibinfo {author} {\bibfnamefont {J.}~\bibnamefont
  {Hindes}}, \bibinfo {author} {\bibfnamefont {K.}~\bibnamefont {Szwaykowska}},
  \ and\ \bibinfo {author} {\bibfnamefont {I.~B.}\ \bibnamefont {Schwartz}},\
  }\href@noop {} {\bibfield  {journal} {\bibinfo  {journal} {Phys. Rev. E}\
  }\textbf {\bibinfo {volume} {94}},\ \bibinfo {pages} {032306} (\bibinfo
  {year} {2016})}\BibitemShut {NoStop}%
\bibitem [{\citenamefont {Osborne}\ and\ \citenamefont
  {Hicks}(2013)}]{Osborne2013}%
  \BibitemOpen
  \bibfield  {author} {\bibinfo {author} {\bibfnamefont {J.}~\bibnamefont
  {Osborne}}\ and\ \bibinfo {author} {\bibfnamefont {G.}~\bibnamefont
  {Hicks}},\ }\href@noop {} {\bibfield  {journal} {\bibinfo  {journal} {Notices
  of the American Mathematical Society}\ }\textbf {\bibinfo {volume} {60}},\
  \bibinfo {pages} {544} (\bibinfo {year} {2013})}\BibitemShut {NoStop}%
\bibitem [{Note1()}]{Note1}%
  \BibitemOpen
  \bibinfo {note} {In practice, we find that randomly generated initial
  conditions are sufficient for computing limit cycles in the single-particle
  system.}\BibitemShut {Stop}%
\bibitem [{\citenamefont {Kuznetsov}(2004)}]{Kuznetsov1}%
  \BibitemOpen
  \bibfield  {author} {\bibinfo {author} {\bibfnamefont {Y.~A.}\ \bibnamefont
  {Kuznetsov}},\ }\href@noop {} {\emph {\bibinfo {title} {Elements of Applied
  Bifurcation Theory}}}\ (\bibinfo  {publisher} {Springer, Berlin},\ \bibinfo
  {year} {2004})\BibitemShut {NoStop}%
\bibitem [{\citenamefont {Strogatz}(2015)}]{StrogatzBook}%
  \BibitemOpen
  \bibfield  {author} {\bibinfo {author} {\bibfnamefont {S.}~\bibnamefont
  {Strogatz}},\ }\href@noop {} {\emph {\bibinfo {title} {Nonlinear Dynamics and
  Chaos: With Applications to Physics, Biology, Chemistry, and Engineering}}}\
  (\bibinfo  {publisher} {Westview Press},\ \bibinfo {year} {2015})\BibitemShut
  {NoStop}%
\bibitem [{\citenamefont {Wiggins}(2003)}]{WigginsBook}%
  \BibitemOpen
  \bibfield  {author} {\bibinfo {author} {\bibfnamefont {S.}~\bibnamefont
  {Wiggins}},\ }\href@noop {} {\emph {\bibinfo {title} {Introduction to Applied
  Nonlinear Dynamical Systems and Chaos}}}\ (\bibinfo  {publisher} {Springer,
  New York},\ \bibinfo {year} {2003})\BibitemShut {NoStop}%
\bibitem [{Note2()}]{Note2}%
  \BibitemOpen
  \bibinfo {note} {Note the agreement between Eq.(16) and the estimate given in
  the first paragraph of Sec.III. The estimate becomes exact by replacing $1$
  with $3/8$. Similarly, if we replace $r$ with the inverse, mean curvature of
  the cylinder, $2\rho$, in Eq.(16), we get an accurate approximation for the
  Hopf bifurcation on the cylinder.}\BibitemShut {Stop}%
\bibitem [{\citenamefont {Moore}(2005)}]{Moore2005FloquetTA}%
  \BibitemOpen
  \bibfield  {author} {\bibinfo {author} {\bibfnamefont {G.}~\bibnamefont
  {Moore}},\ }\href@noop {} {\bibfield  {journal} {\bibinfo  {journal} {SIAM J.
  Numer. Anal.}\ }\textbf {\bibinfo {volume} {42}},\ \bibinfo {pages} {2522}
  (\bibinfo {year} {2005})}\BibitemShut {NoStop}%
\end{thebibliography}%

\end{document}